\documentclass[pdflatex,sn-basic,iicol]{sn-jnl}

\usepackage{graphicx}%
\usepackage{multirow}%
\usepackage{amsmath,amssymb,amsfonts}%
\usepackage{amsthm}%
\usepackage{mathrsfs}%
\usepackage[title]{appendix}%
\usepackage{xcolor}%
\usepackage{textcomp}%
\usepackage{manyfoot}%
\usepackage{booktabs}%
\usepackage{algorithm}%
\usepackage{algorithmicx}%
\usepackage{algpseudocode}%
\usepackage{listings}%

\theoremstyle{thmstyleone}%
\theoremstyle{thmstyletwo}%

\theoremstyle{thmstylethree}%

\begin{document}

\title[Comprehensive VLBI observations of Galileo satellites with the AuScope array]{Comprehensive VLBI observations of Galileo satellites with the AuScope array}


\author*[1]{\fnm{David} \sur{Schunck}}\email{david.schunck@utas.edu.au}

\author[1]{\fnm{Lucia} \sur{McCallum}}\email{lucia.mccallum@utas.edu.au}

\author[1]{\fnm{Jamie} \sur{McCallum}}\email{jamie.mccallum@utas.edu.au}

\author[1]{\fnm{Tiege} \sur{McCarthy}}\email{tiegem@utas.edu.au}

\affil[1]{\orgdiv{School of Natural Sciences}, \orgname{University of Tasmania}, \orgaddress{\street{PO Box 807}, \city{Hobart}, \postcode{7004}, \state{Tasmania}, \country{Australia}}}



\abstract{Interest in the topic of geodetic co-location in space and space ties has recently intensified within the geodetic community, particularly following the approval of the European Space Agency’s (ESA) Genesis mission. From the perspective of Very Long Baseline Interferometry (VLBI), observations of Earth-orbiting satellites are not standard practice yet. To enable VLBI support for future co-location satellite missions, such observations must be integrated into the VLBI processing chain. In this study, we present comprehensive VLBI observations of Galileo navigation satellites conducted with the Australian AuScope VLBI array. Using the 12-m antennas in Hobart, Katherine and Yarragadee equipped with VLBI Global Observing System (VGOS) instrumentation, Galileo E1 and E6 signals were observed in test experiments and a series of four full-scale 24-hour observing sessions. We present the estimation of VLBI station coordinates from observations to navigation satellites,
thereby demonstrating, for the first time, inter-technique ties between the VLBI and Global Navigation Satellite System (GNSS) frame. We describe the processing strategy, including correlation, fringe fitting, precision assessment and satellite tracking approach. Delay observables achieve precisions of a few picoseconds in the E1 band and several tens of picoseconds in the E6 band for 1-s integration times. However, unmodelled signals on the order of several hundred picoseconds are found in the residual delays. Estimated station coordinates agree with a priori values at the metre level, while baseline lengths agree at the sub-metre level. These results demonstrate the feasibility of large-scale VLBI observations to GNSS satellites and provide critical groundwork for future co-location satellite missions such as Genesis.}


\maketitle

\section{Introduction}\label{sec:01_introduction}

An accurate and stable terrestrial reference frame is a fundamental requirement for geodesy and Earth sciences, underpinning applications ranging from precise orbit determination and Earth rotation studies to sea-level monitoring and plate tectonics \citep{GGCE2024}. The latest iteration of the International Terrestrial Reference Frame (ITRF), the ITRF2020 \citep{Altamimi2023jog}, is currently the most accurate global realisation of a terrestrial reference system. It is derived from the combination of four space-geodetic techniques: Very Long Baseline Interferometry (VLBI), Global Navigation Satellite Systems (GNSS), Satellite Laser Ranging (SLR) and Doppler Orbitography and Radiopositioning Integrated by Satellite (DORIS). Despite continuous improvements, the long-term accuracy and stability of the ITRF remain at the level of several millimetres, which is insufficient to meet the Global Geodetic Observing System (GGOS) target of 1~mm accuracy and 0.1~mm/year stability \citep{Plag2009ggos}.

A major limitation in the current realisation of the ITRF arises from the linking of the individual techniques. Inter-technique ties are established through terrestrial local tie surveys at co-location sites \citep[e.g.][]{Matsumoto2022EPS}. However, discrepancies of up to several centimetres between local ties and global space geodetic estimates are detected \citep{Altamimi2023jog}, limiting the overall consistency of the combined frame. A promising approach to overcome this shortcoming is the concept of co-location applied to satellites, so-called ``space ties", whereby an Earth-orbiting satellite is equipped with instruments of multiple space geodetic techniques \citep{Rothacher2009ggos}. Space ties offer the potential to establish globally consistent, observation-level ties between techniques and to independently validate ground-based local tie surveys. The importance of this concept was recognised by the research community and mission proposals, such as GRASP \citep{BarSever2009cospar} and EGRASP/Eratosthenes \citep{Biancale2016egrasp}, were submitted, but not funded in the past.

With the approval of the European Space Agency’s (ESA) Genesis mission in November 2022, the interest in VLBI observations to satellites increased again. The Genesis satellite will be the first co-location satellite combining the instruments of all four space geodetic techniques on a single satellite platform \citep{Delva2023genesis}, enabling in-orbit space ties for future realisations of the ITRF. Scheduled for launch in 2028, the research community is now called upon to implement the mission in the processing chains of the space geodetic techniques and successfully support it with observations. For the VLBI technique, these efforts are organised through the International VLBI Service for Geodesy and Astrometry \citep[IVS,][]{Nothnagel2017}, specifically working group~7 on satellite observations with VLBI \footnote{\url{https://ivscc.gsfc.nasa.gov/about/wg/wg7/index.html}}.

Previous simulation studies underscore the theoretical potential of VLBI observations to Earth-orbiting satellites, both GNSS constellations and dedicated co-location satellites, for the establishment of ties to other space geodetic techniques \citep[e.g.][]{Plank2014jog, Anderson2018jog, Klopotek2020jog, Pollet2023, Wolf2023eps, Schunck2024jgs, Schunck2024rs}. \citet{Boehm2024iag} summarise the opportunities with upcoming satellites carrying VLBI transmitters. However, their realisation is a significant challenge as such observations are fundamentally different from regular VLBI operations of the IVS. While the general feasibility has been demonstrated in the past, observations are non-standard and far from routine. In several studies, test experiments were performed to observe and correlate the signals of GNSS satellites \citep{Tornatore2014iag, Haas2014ivs, Haas2017evga, Plank2017jog, Schunck2024iag}. Understood as pathfinders, they revealed some of the issues inherent to VLBI observations of artificial signals of near-field targets. In further studies, antennas of the IVS network were employed to observe the Chinese Chang'e~3 lunar lander \citep{Klopotek2019eps} and Chang'e~5 lunar orbiter \citep{Gan2025rs}, and observations with the AuScope VLBI array were performed to the APOD mission \citep{Hellerschmied2018sen, Sun2018asr}. \citet{Skeens2023} implemented a slightly different observing concept, in which an interferometric response is derived from observations of a VLBI radio telescope and modified geodetic GNSS antenna.

More practical observations are required to optimise the technique and work towards a full implementation into routine operations and procedures. Many aspects inherent to VLBI observations to satellites, such as tracking, recording setup, correlation or the support in analysis software, remain unresolved and require further development. Although several GNSS constellations are in operation at present, offering the opportunity for practical observations, their signals differ significantly to that of natural geodetic radio sources. Geodetic VLBI traditionally observes weak, broadband radio emission from distant extra-galactic sources at S/X bands or, more recently, across wide frequency ranges between 3~and 13.5~GHz using the VLBI Global Observing System (VGOS). In contrast, GNSS satellites transmit extremely strong, narrowband signals in L-band with many orders of magnitude more power. This intrinsic mismatch in signal characteristics and frequency allocation has prevented routine VLBI observations of GNSS satellites and hindered a direct observational link between the two techniques. As a result, the practical feasibility, limitations in the processing chain and achievable precision remain insufficiently understood.

In recent works, we demonstrated the capability of the AuScope VLBI array to perform observations of GNSS constellations in L-band \citep{McCallum2025pasp, Schunck2025phd}. Using the three VGOS telescopes in Hobart, Katherine and Yarragadee, a set of Global Positioning System (GPS) satellites were observed over several hours. The observations were processed through the standard VLBI processing chain, deriving geodetic baseline delays. This capability was enabled through the implementation of an additional signal chain for L-band, from the same receiver feed.

In this work, we present extensive VLBI observations of Galileo satellites using the radio telescopes of the Australian AuScope VLBI array. Building upon the work by \cite{McCallum2025pasp} and \cite{Schunck2025phd}, this study utilises the newly established capability of the antennas in Hobart, Katherine and Yarragadee to receive and process navigation signals in L-band. Besides two test experiments, four full-scale 24-h VLBI observing sessions were conducted, representing a significant expansion in scale of VLBI observations to satellites compared to previously presented experiments in literature. The scope of observations is of unprecedented comprehensiveness.
Using standard 12-m VGOS antennas and the conventional VLBI processing chain, we demonstrate the derivation of geodetic delay observables in a routine process. For the first time, we derive station position estimates from the analysis of Galileo E1 and E6 signals. These practical observations represent a first realisation of inter-technique ties between VLBI and GNSS and provide important insights into the achievable precision and systematic effects.

\subsection{Objectives of the study}

The aim of this work is to investigate the data recording, processing and analysis of VLBI observations of satellites of the Galileo constellation with the VGOS antennas of the AuScope VLBI array.
In contrast to natural geodetic radio sources, the signals of GNSS satellites are deterministic and modulated. In VLBI, the incoming signals are typically digitised as 2-bit samples. A goal of the study is to investigate and quantify the potential benefit of a higher digitisation bit depth of 8-bit. This is done by analysing the precision of the observables and comparing post-fit residuals of a geodetic analysis. As the employed antennas are not yet capable to continuously track a satellite, we further investigate the impact of a stepwise tracking approach on the results. Based on the four 24-h observing sessions, we aim to demonstrate the feasibility of estimating VLBI antenna coordinates in the satellite frame.

Section~\ref{sec:02_observations} describes the observational setup, including the antenna network, scheduling, satellite tracking, L-band signal chains, data recording, setup of interferometric delay models, correlation, fringe fitting, post-processing and analysis. Section~\ref{sec:03_overview_experiments} provides an overview of the experiments which were conducted in the scope of this work. The results of the study objectives are presented in Sect.~\ref{sec:04_results}. Section~\ref{sec:05_dicussion} discusses the work, and Sect.~\ref{sec:06_outlook} provides an outlook to future investigations.

\section{Observations}\label{sec:02_observations}

\subsection{Antenna network}\label{sec:02_antenna_network}

\begin{figure}[t!]
	\centering
	\includegraphics[width=\linewidth]{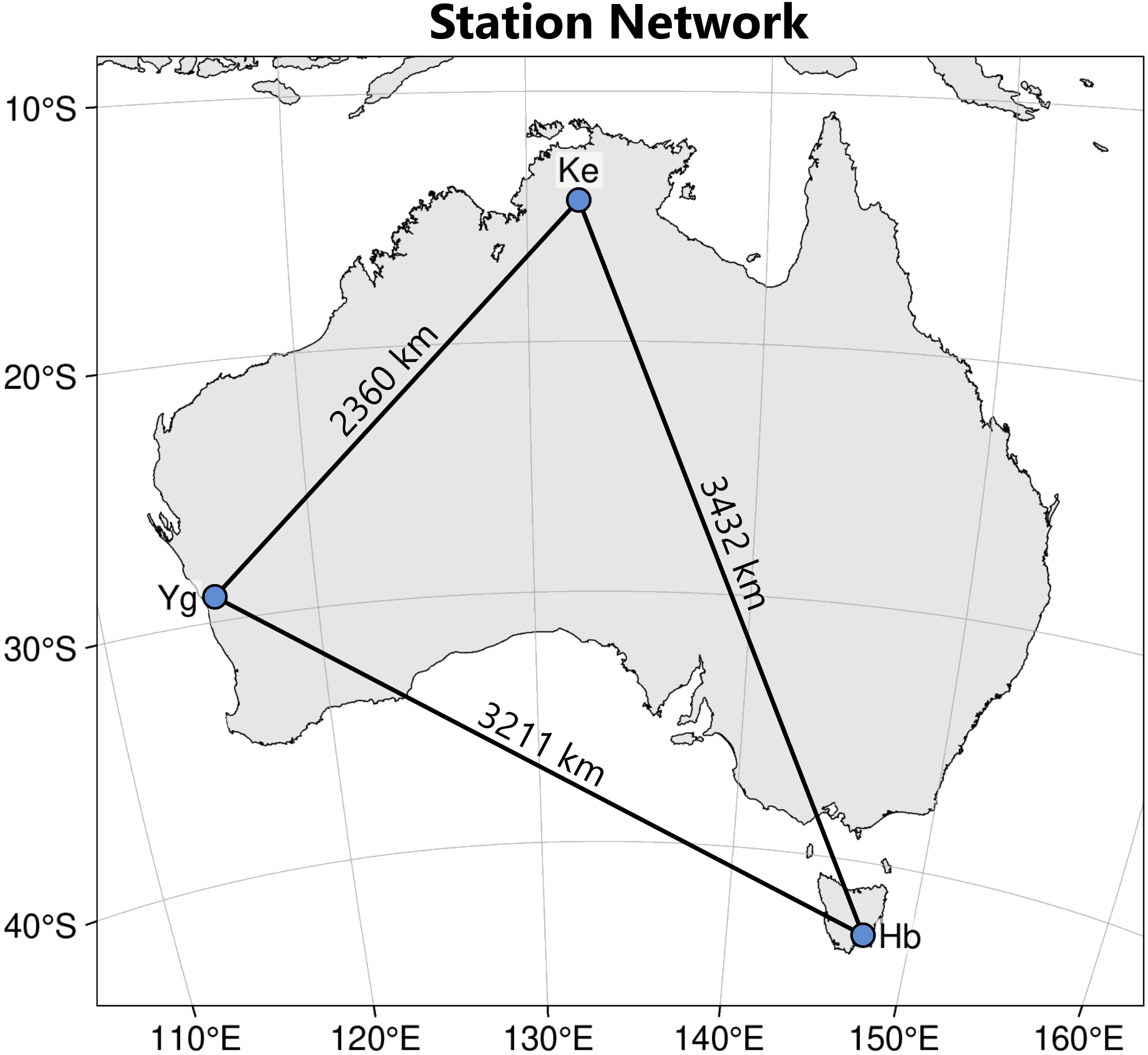}
	\caption{Locations of the antennas used in the experiments of this work in Hobart (Hb), Katherine (Ke) and Yarragadee (Yg).}
	\label{fig:station_overview}
\end{figure}

The experiments presented in this paper were performed with the three 12-m antennas in Hobart (Hb), Katherine (Ke) and Yarragadee (Yg) of the AuScope VLBI array. The antennas have an azimuth/elevation (AzEl) mount with a slew speed of 300$^\circ/$min in azimuth and 75$^\circ/$min in elevation.
The telescopes are owned and operated by the University of Tasmania, with operational IVS observing activities funded through Geoscience Australia. Technical specifications of the telescopes and receiver design are detailed by \cite{Lovell2013jog} and \cite{McCallum2022jog}.
The antenna locations are shown in Fig.~\ref{fig:station_overview}.

\subsection{Scheduling}\label{sec:02_scheduling}

\begin{figure}[t!]
	\centering
	\includegraphics[width=\linewidth]{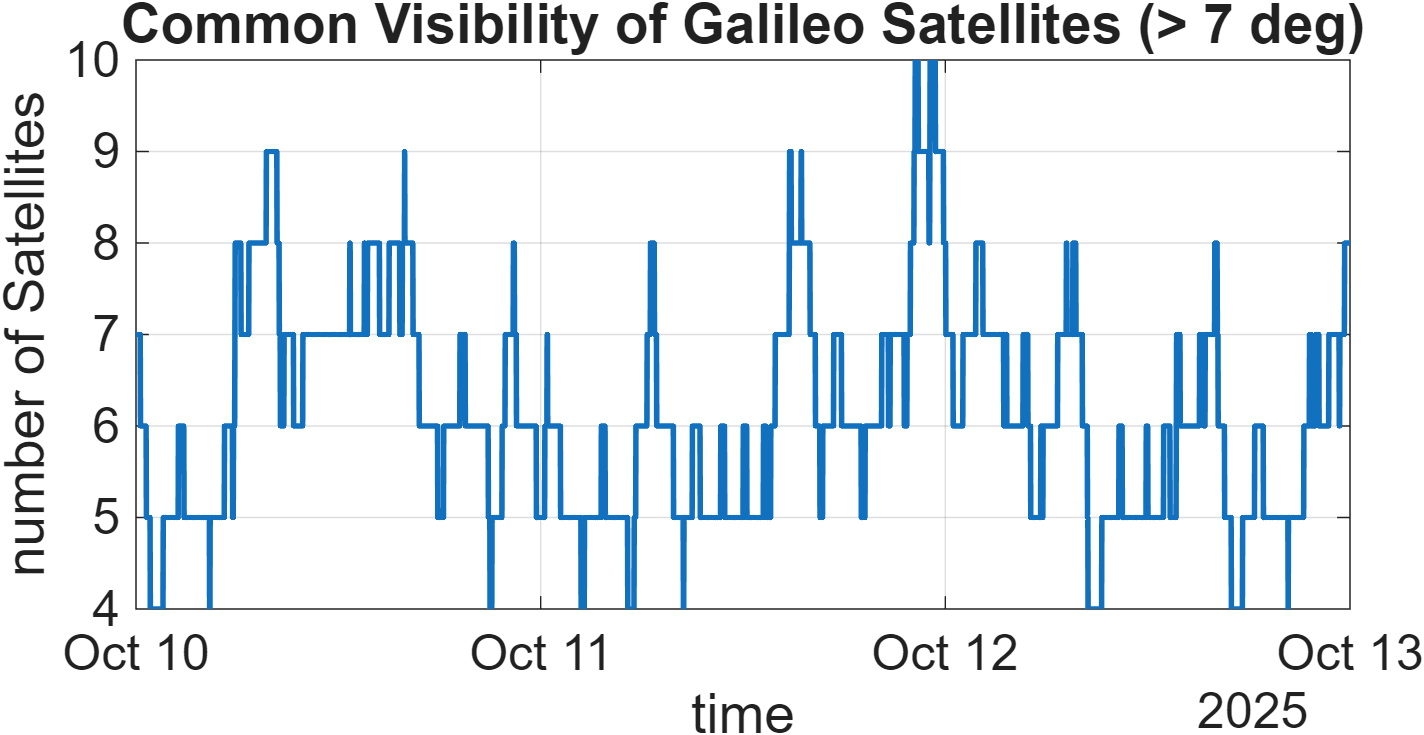}
	\caption{Common visibility of Galileo satellites at elevations larger than 7$^{\circ}$ from the antennas in Hobart, Katherine and Yarragadee over three days.}
	\label{fig:common_vis}
\end{figure}

In geodetic VLBI, a schedule is the optimised observation plan that coordinates which antennas simultaneously observe which sources at what times. In contrast to schedules which are optimised for global antenna networks and employ source catalogues of several hundred sources, the scheduling process for the experiments in this work is much simpler.
This is because the network of stations is limited to three antennas and because the number of sources is limited to only about 30~satellites of the Galileo constellation. The number of satellites which are visible from all stations simultaneously is significantly less than the number of sources in a typical source catalogue. Figure~\ref{fig:common_vis} visualises the number of Galileo satellites visible from Hobart, Katherine and Yarragadee at the same time over a period of three days in October 2025. A minimum of 4~and maximum of 10~satellites were visible simultaneously at elevations above 7$^{\circ}$.

The two short test experiments conducted in the scope of this work, each comprising four scans, were scheduled manually. In contrast, the four full-scale 24-h observing sessions were scheduled automatically using the state-of-the-art scheduling software VieSched++ \citep{Schartner2019viesched}.
All satellites of the Galileo constellation that were classified as operational by the European GNSS Service Centre\footnote{\url{https://www.gsc-europa.eu/system-service-status/constellation-information}} (GSC) at the time of scheduling were included. Satellite orbit information was provided to the scheduling software in the form of two-line elements\footnote{\url{https://celestrak.org/NORAD/documentation/tle-fmt.php}} (TLEs).
The resulting schedules were available in the standard experiment formats \texttt{VEX} and \texttt{skd}. However, in their current versions, neither format preserves the satellite orbit information supplied to the scheduler or offers mechanisms to reference external orbit data. Instead, each scan represents the satellite by a single fixed position in the celestial frame, analogous to a stationary natural geodetic radio source.

In standard geodetic VLBI observing sessions, schedules are provided to the antennas through the NASA Field System\footnote{\url{https://github.com/nvi-inc/fs}} as SNAP files. These files contain commands for the Field System to govern antenna movement, recording management and logging of antenna information.
SNAP files are typically generated from the schedule files using the \textit{drudg} program. In its current version it does not support moving satellites as observation targets and is not able to write dedicated Field System commands for satellite tracking. To generate SNAP files containing satellite information, a custom script was developed to merge the schedule information from the \texttt{VEX} file with satellite orbit data from TLEs, producing antenna-specific SNAP files suitable for the presented experiments.

\subsection{Satellite tracking}\label{sec:02_tracking}

To ensure a reliable signal acquisition, a satellite must remain within the antenna's field of view throughout an observation. While the data is recorded, the moving target should ideally remain centred within the antenna main beam to minimise amplitude variations in the received signal \citep{McCallum2017jog, Hellerschmied2018phd}. Hence, it is desired that the antenna slews smoothly and constantly while tracking, adopting the antenna's slew speed on the satellite motion without alternating acceleration and deceleration of the antenna axes. This approach is often referred to as continuous tracking.

The Field System supports satellite tracking by being able to process TLE orbit data and providing commands specifically for satellite tracking \citep{Himwich2010ivs}. However, the implementation of continuous tracking with the Field System is station-dependent as it requires special programming of the antenna interface. It has to be realised according to the interface specifications and requirements of the antenna control unit (ACU). Some antennas' ACU may not be able to support this mode.

\begin{figure}[t!]
	\centering
	\includegraphics[width=\linewidth]{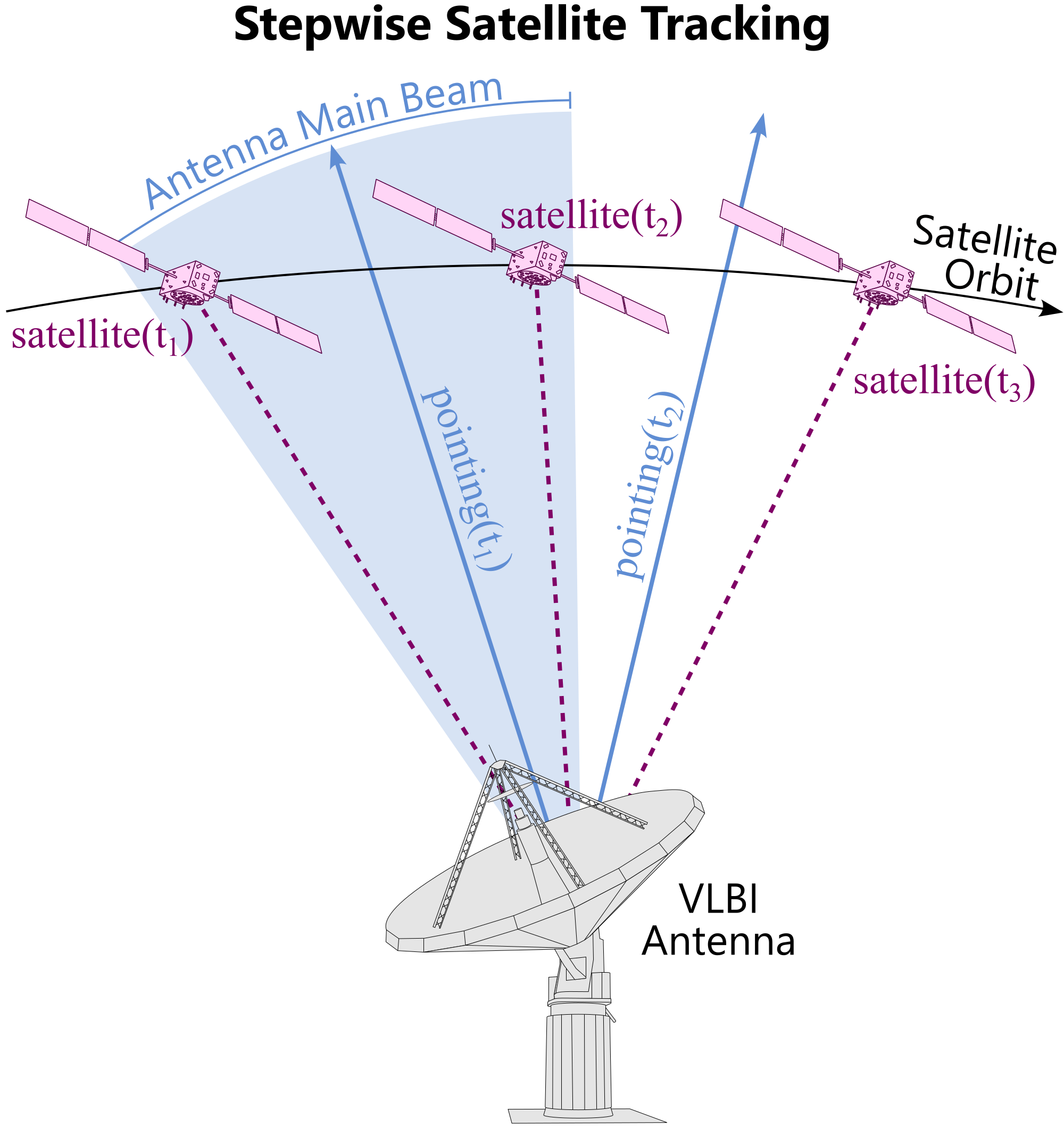}
	\caption{Illustration of the stepwise satellite tracking approach employed in the experiments of this work. At time epoch $t_i$, the antenna points ahead of the satellite in along-track direction at the midpoint of the orbit arc between the time epochs $t_i$ and $t_{i+1}$ to reduce the maximum angular separation between the topocentric satellite direction and the antenna pointing.}
	\label{fig:satellite_tracking}
\end{figure}

The antennas used in this study are not yet capable of continuously tracking satellites across the local sky \citep{McCallum2025pasp, Schunck2024iag, Schunck2025phd}. Hence, a stepwise tracking approach is employed, in which the antenna pointing is updated at fixed time intervals in order to keep the moving satellite within the antenna's field of view during the scans. In doing so, the satellite track is essentially treated as a sequence of sources with static coordinates. The satellite tracking approach is illustrated in Fig.~\ref{fig:satellite_tracking}.
For the observations presented in this work, we used the \textit{satellite} command of the Field System, which accepts TLEs as input for satellite information. The Field System propagates the satellite positions using the SGP4 propagator within the \textit{predict} program. The \textit{satellite} command points the antenna to the evaluated satellite position at the time of the command. It can be re-issued at regular intervals to follow the satellite in discrete steps. To reduce the maximum angular separation between the nominal topocentric satellite position and the antenna pointing at the end of each reposition interval, we additionally used the Field System's \textit{satoff} command. The command applies an along-track offset, effectively pointing the antenna ahead of the satellite. We set the offset to half the \textit{satellite} command's re-command interval so that at the time of each commanding epoch the antenna points to the midpoint of each satellite orbit arc between successive pointing updates.

A disadvantage of stepwise tracking is that the satellite cannot remain continuously centred in the antenna main beam. Because the antenna updates its pointing only at discrete intervals while the satellite moves continuously, the target periodically drifts off-centre, leading to time-dependent amplitude fluctuations and potentially systematic phase effects. The severity of these impacts depends on the satellite orbit height, apparent motion, antenna beam width and the chosen update interval. In this work, we tested and compared reposition intervals of 30, 20, 10 and 5~s. Attempts to apply a stepwise tracking with a 1- or 2-s reposition interval failed, because the individual positioning commands took too much time to be processed by the Field System. The high rate of commands overloaded the Field System, and it became unresponsive as also reported by \citet{Hellerschmied2018phd}.

\subsection{L-band signal chain}

The three antennas of the AuScope VLBI array have QRFH VGOS feed horns \citep{Akgiray2013vgos} installed, which are designed to operate in the frequency range between 2~and 14~GHz to receive signals from faint celestial sources in S-band and the VGOS frequencies. Signals with frequencies below 2~GHz experience a strong attenuation. As the navigation signals of GNSS are several magnitudes stronger than the signals of natural geodetic radio sources, the attenuation is favourable to protect the antenna back ends in the control room from oversaturation and damage.

\begin{figure}
	\centering
	\includegraphics[width=\linewidth]{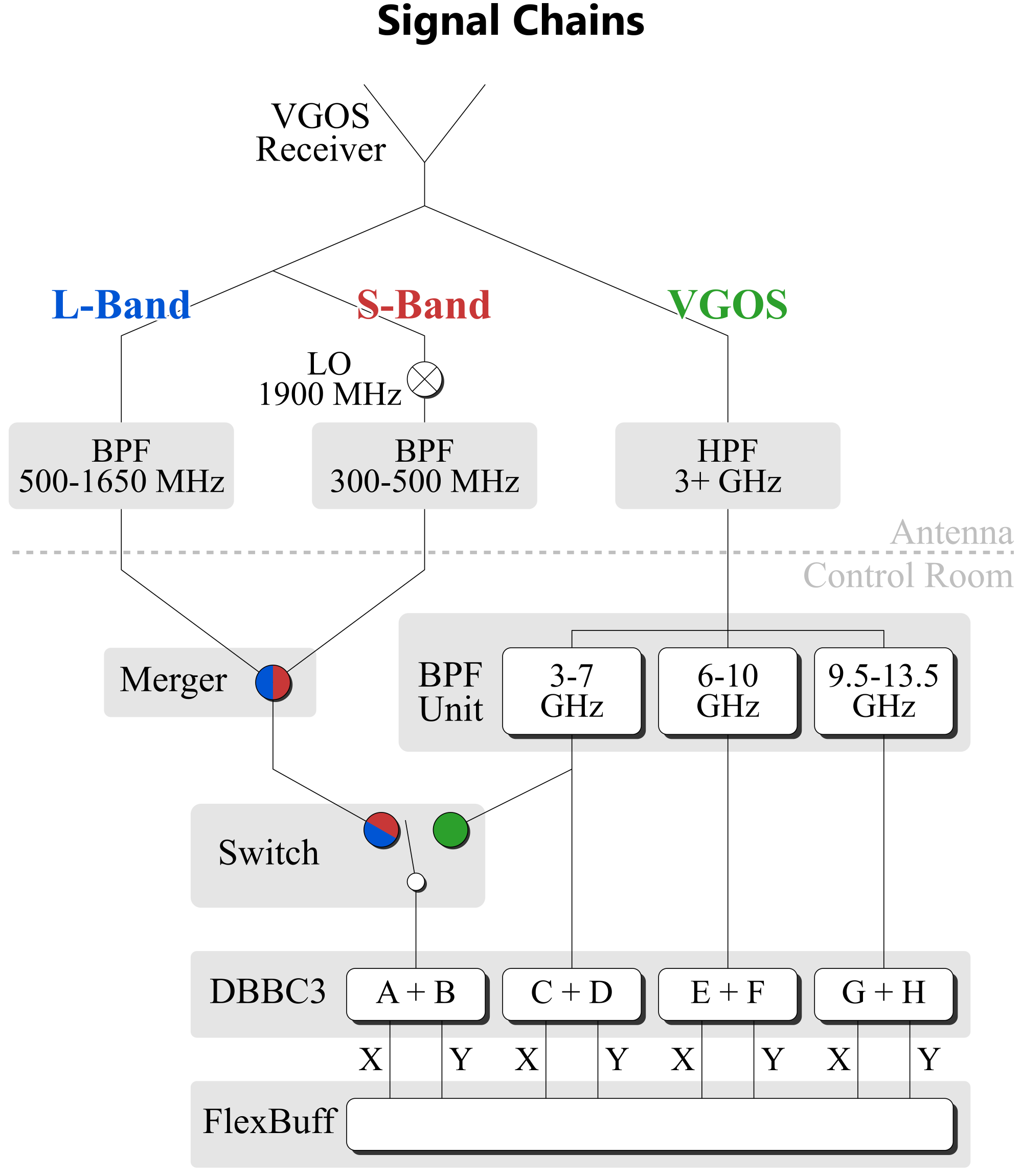}
	\caption{Simplified illustration of the L-band, S-band and VGOS signal chains at the 12-m VLBI antennas in Hobart, Katherine and Yarragadee.}
	\label{fig:Lband_signal_chain}
\end{figure}

The capability to record signals in L-band is newly accessible with the antennas of the AuScope VLBI array \citep{McCallum2025pasp, Schunck2025phd}. The diagram in Fig.~\ref{fig:Lband_signal_chain} shows the signal chains for L-band, S-band and the VGOS frequencies. The received signal is amplified by a VGOS cryogenic low noise amplifier (LNA) before it is split. In the L-band signal chain, the signal passes through a second-stage LNA and is filtered between 500~and 1650~MHz. The S-band signal is mixed with a local oscillator with a frequency of 1900~MHz and filtered between 300~and 500~MHz, the equivalent of 2200~and 2400~MHz at sky frequency. 
For the Hobart station, a stricter upper limit of 400~MHz is applied due to local radio-frequency interference (RFI).
The L- and down-converted S-band signals travel via coaxial cables to the control room, where they are merged. A signal switch which can be operated remotely can either forward the bandpass filtered signal from the VGOS signal chain ($3$~to $7$~GHz) or the combined signal from the L- and S-band to the samplers A and B of the analog-to-digital converter DBBC3 \citep{Tuccari2014ivs}. As VGOS feed horns are linearly polarised, the signal chains in front of the DBBC3 shown in Fig.~\ref{fig:Lband_signal_chain} are installed twice for both the X and Y polarisation. Sampler~A digitises the X polarisation, while sampler~B digitises the Y polarisation. A FlexBuff recorder writes the data to disks. It should be noted that tests have shown that the sensitivity in S-band is not affected by merging the S-band with the L-band signal for observations of faint natural geodetic radio sources.

\begin{figure*}[t!]
	\centering
	\includegraphics[width=1\linewidth]{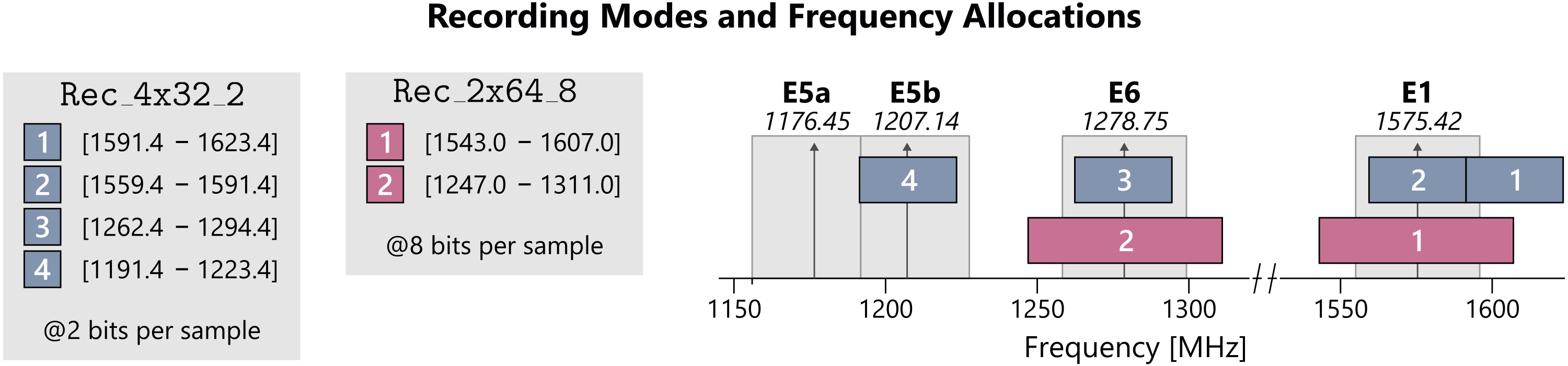}
	\caption{Overview of the applied recording modes \texttt{Rec\_4x32\_2} and \texttt{Rec\_2x64\_8} (left) and the allocated frequency channels to the Galileo frequency bands (right). Information is given about the channel numbers, exact frequency range, bit-depth and corresponding Galileo frequency bands. The vertical upwards pointing arrows represent the respective band centre frequencies.}
	\label{fig:recording_mode_frequencies}
\end{figure*}

\subsection{Data recording}\label{sec:02_data_recording}

Two data recording modes were applied in different experiments of this study.
Recording mode \texttt{Rec\_4x32\_2} records the satellite signals in four channels with a width of 32~MHz each. The samples are digitised with a bit-depth of~2, which is the common standard for geodetic VLBI observations for the signals of natural geodetic radio sources. The exact frequency ranges of the individual channels and the allocation with respect to the Galileo frequency bands are presented in Fig.~\ref{fig:recording_mode_frequencies}. The channels 4, 3 and 2 cover the E5b, E6 and E1 Galileo navigation signals, respectively. Additionally, channel~1 is positioned at the upper L-band, edge-to-edge at the upper bound of channel~2, with the aim of detecting fringes of natural radio sources.
The Galileo E5a frequency band was not covered by a recording channel because test experiments have shown that a channel in the band shows very weak fringe detections with the present network. This is because the sensitivity of the VGOS feed horns sharply declines towards the lower range of L-band.
While the emitted satellite signal is circularly polarised, the antennas of the AuScope VLBI array are equipped with linearly polarised receivers. Both polarisations were recorded. Each frequency channel is then recorded twice, once in X polarisation and a second time in Y polarisation.
Overall, mode \texttt{Rec\_4x32\_2} has a data recording rate of 1~Gbps.

Recording mode \texttt{Rec\_2x64\_8} records the satellite signals in two channels with a width of 64~MHz each. The samples are digitised with 8~bits. As discussed by \cite{McCallum2017jog}, the higher bit depth leads to digitised samples of the signal which better represent the modulated, deterministic signals that are transmitted by navigation satellites like Galileo. This has the potential to be preferable over the standard 2-bit digitisation in the correlation process. The exact frequency ranges of the individual channels and the allocation with respect to the Galileo frequency bands are presented in Fig.~\ref{fig:recording_mode_frequencies}. The channels 1 and 2 cover the E1 and E6 Galileo navigation signals, respectively. This mode only focuses on the two Galileo bands for which strong fringes can be detected. Due to the increased bit-depth, the overall recording rate of mode \texttt{Rec\_2x64\_8} equals 4~Gbps.

\begin{figure*}
	\centering
	\includegraphics[width=.8\linewidth]{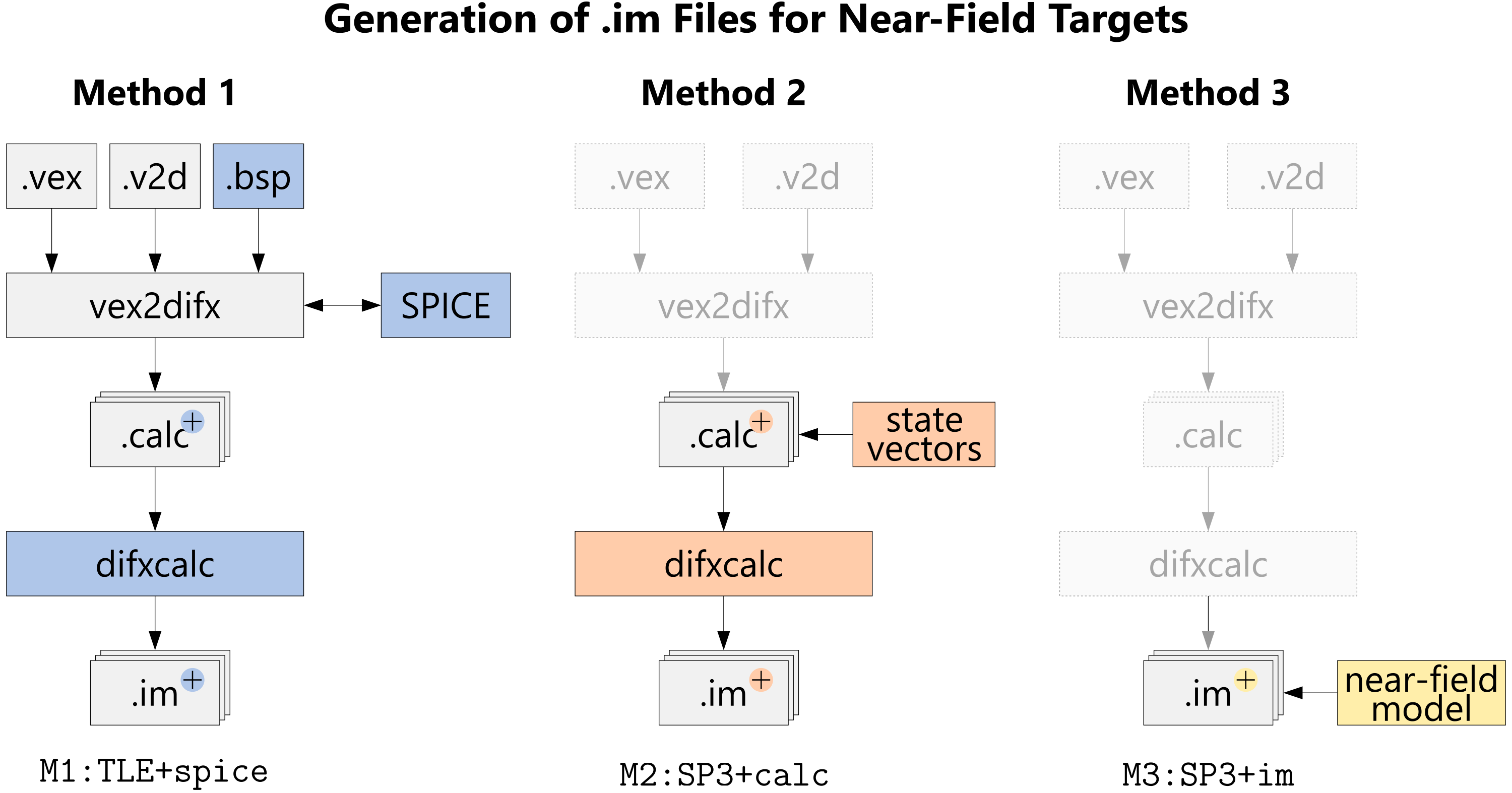}
	\caption{Three methods to include orbit information of near-field targets in interferometric model file sets (\texttt{.im} files) in the workflow of the DiFX software correlator. In this work, we use each method, in the following referred to as \texttt{M1:TLE+spice}, \texttt{M2:SP3+calc} and \texttt{M3:SP3+im}.}
	\label{fig:im_workflow}
\end{figure*}

\subsection{Interferometric models to the correlator for near-field sources}\label{sec:02_interferometric_models}

The data were correlated using the DiFX software correlator \citep{Deller2007pasp, Deller2011pasp}. The correlator requires so-called interferometric input models which represent the expected, modelled delays as function of time. Accurate a priori models are crucial for the correlator to align the data streams recorded at each station and retain coherence throughout the duration of the scan. In the DiFX workflow, the interferometric models are contained in the \texttt{.im} files.

In contrast to far-field observations, interferometric models of near-field sources must incorporate accurate information about the target's orbit.
The typical DiFX workflow used for operational geodetic VLBI experiments can be adjusted in several ways to allow the inclusion of satellite orbit information in the \texttt{.im} files. At present, there are three methods, which are illustrated in Fig.~\ref{fig:im_workflow}. In general, DiFX requires a combined schedule file in \texttt{VEX} format (\texttt{.vex} file) for information for the configuration of the experiment including frequency set-ups, antenna scheduling and source information as well as a \texttt{.v2d} file which is used to specify correlation options, set a priori clock parameters, set Earth Orientation Parameters (EOPs) and define data locations on the computing cluster.

In method~1 (Fig.~\ref{fig:im_workflow} left), the satellite orbit information enters the DiFX workflow in the form of a binary SPICE kernel file. This \texttt{.bsp} file contains orbits as orbital parameters or state vectors as a function of time for one or more Earth-orbiting satellites. They are generated with the SPICE toolkit\footnote{\url{https://naif.jpl.nasa.gov/naif/toolkit.html}}. The program \texttt{vex2difx} uses a program of the SPICE library to propagate and interpolate the orbits at the time of the scans. The evaluated satellite positions are written in \texttt{.calc} files as state vector tables. The program \texttt{difxcalc} has the capability to use the state vector entries to generate interferometric model files (\texttt{.im} files) for satellites by applying a near-field delay model \citep{Gordon2016ivs}. Three delay models are implemented. We use the Duev model \citep{Duev2012}. In the \texttt{.im} files, the delay models are described as 5th-order polynomials each valid for 2~minutes. In this work, we utilise this method to generate interferometric model files by providing a \texttt{.bsp} file derived from TLEs. In the following, this model is referred to as \texttt{M1:TLE+spice}.

In method~2 (Fig.~\ref{fig:im_workflow} centre), \texttt{vex2difx} is not provided with an orbit file.
Instead, the satellite orbits are propagated and interpolated at the times of the scans to derive satellite state vectors externally. These are directly written into the \texttt{.calc} files, extending them by the satellite position and velocity.
The implemented Duev near-field delay model in \texttt{difxcalc} is applied to generate \texttt{.im} files for the satellite scan. In this work, we utilise this method by generating state vectors derived from SP3 files. In the following, this model is referred to as \texttt{M2:SP3+calc}.

In method~3 (Fig.~\ref{fig:im_workflow} right), \texttt{.im} files are generated externally, not making use of the near-field support in the DiFX workflow.
As described in \citet{McCallum2017jog} and \citet{Hellerschmied2018phd}, we use the Vienna VLBI and Satellite software \citep[VieVS,][]{Boehm2018pasp} to generate computed delays. 
VieVS has the Klioner near-field delay model for Earth satellites \citep{Klioner1991agu} implemented. It is employed to derive geocentric delays in 1-s intervals. Based on these discrete values, 5th-degree polynomials are estimated and their coefficients are written in the \texttt{.im} files.
In this work, we utilise this method by evaluating SP3 files with VieVS. In the following, this model is referred to as \texttt{M3:SP3+im}

In this section, we describe three approaches for incorporating satellite orbit data into the DiFX workflow. For each method, we provide a representative implementation. Consequently, the subsequent comparison of the three resulting sets of \texttt{.im} files reflects not only differences in the underlying methods and theoretical models, but also the specific implementations adopted for each approach.

\subsection{Correlation and fringe fitting}\label{sec:02_correlation}

\begin{table*}[t!]
	\centering
	\begin{tabular*}{\linewidth}{@{\extracolsep{\fill}} l|cccc}
		\toprule
		Fringe Mode                & Rec Mode              & Bandpass & Fringe Freq E1      & Fringe Freq E6 \\ \midrule
		\texttt{Frg\_4x32\_2\_32}  & \texttt{Rec\_4x32\_2} & full    & [1559.40 - 1591.40] & [1262.40 - 1294.40] \\ \midrule
		\texttt{Frg\_2x64\_8\_32}  & \texttt{Rec\_2x64\_8} & reduced & [1559.40 - 1591.40] & [1262.40 - 1294.40] \\
		\texttt{Frg\_2x64\_8\_64}  & \texttt{Rec\_2x64\_8} & full    & [1543.00 - 1607.00] & [1247.00 - 1311.00] \\
		\texttt{Frg\_2x64\_8\_Opt} & \texttt{Rec\_2x64\_8} & optimised & [1558.42 - 1592.42] & [1265.75 - 1291.75] \\
		\bottomrule
	\end{tabular*}
	\caption{Overview of the employed fringe fitting modes with their respective frequency ranges in the Galileo E1 and E6 frequency bands employed in \texttt{fourfit}.}
	\label{tab:fringe_modes}
\end{table*}

The correlation of the observations was carried out on a computing cluster running DiFX version 2.6.2, based at the Mt Pleasant observatory near Hobart. It was performed with a uniform spectral resolution of 125~kHz (256~channels over 32~MHz) and accumulation periods of 1~s.
The correlation results were converted to files of the \texttt{mark4} format. The fringe fitting was carried out with \texttt{fourfit}, which is part of the Haystack Observatory Postprocessing System (HOPS\footnote{\url{http://www.haystack.mit.edu/tech/vlbi/hops.html}}). Used as common standard in geodetic VLBI, \texttt{fourfit} provides total delays referenced to reception at the first station at integer second time. The fringe fitting was performed in single-band mode, meaning each channel was fringe fitted individually. As the data was recorded in two linear polarisations, we consider the individual polarisation products XX, YY, XY and YX. A phase and delay calibration as, for example, described in \cite{Gan2025rs} was not realised. This is because the antenna's phase calibration system is not functional in the L-band signal chain.
Furthermore, a manual phase calibration using the signals of natural geodetic radio sources is not feasible either. Although, channels~1 and~2 of recording mode \texttt{Rec\_4x32\_2} and channel~1 of recording mode \texttt{Rec\_2x64\_8} were positioned at the most sensitive section of the L-band signal chain, fringes to strong natural geodetic radio sources were not detected among the 12-m antennas.

In the fringe fitting process, the frequency width of the recorded channels were partly reduced to smaller frequency ranges. This is in the following referred to as \textit{Fringe Mode}. All applied fringe modes are presented in Table~\ref{tab:fringe_modes}. Fringe mode \texttt{Frg\_4x32\_2\_32} is applied for the data recorded in mode \texttt{Rec\_4x32\_2} and utilises the entire width of 32~MHz of the channels in the E1 and E6 bands for fringe fitting. For the data recorded in mode \texttt{Rec\_2x64\_8} three different fringe modes are applied. Mode \texttt{Frg\_2x64\_8\_32} limits the frequency ranges in the channels for E1 and E6 to a width of 32~MHz. As the frequency ranges are the same as for the fringe mode \texttt{Frg\_4x32\_2\_32}, the 8- and 2-bit data can be compared employing the same frequencies with the only difference being the sample bit depth. The fringe mode \texttt{Frg\_2x64\_8\_64} does not reduce the channel widths and utilises the entire recorded frequency range of 64~MHz for fringe fitting. Lastly, for the fringe mode \texttt{Frg\_2x64\_8\_Opt}, the frequency ranges for the channels were tuned empirically around the respective centre frequencies of the E1 and E6 bands at 1575.42~MHz and 1278.75~MHz respectively to optimise the resulting signal-to-noise ratio (SNR). In this fringe mode, the channel bandpass for E1 and E6 are reduced to 34~MHz and 26~MHz, respectively.

\begin{figure*}
	\centering
	\includegraphics[width=\linewidth]{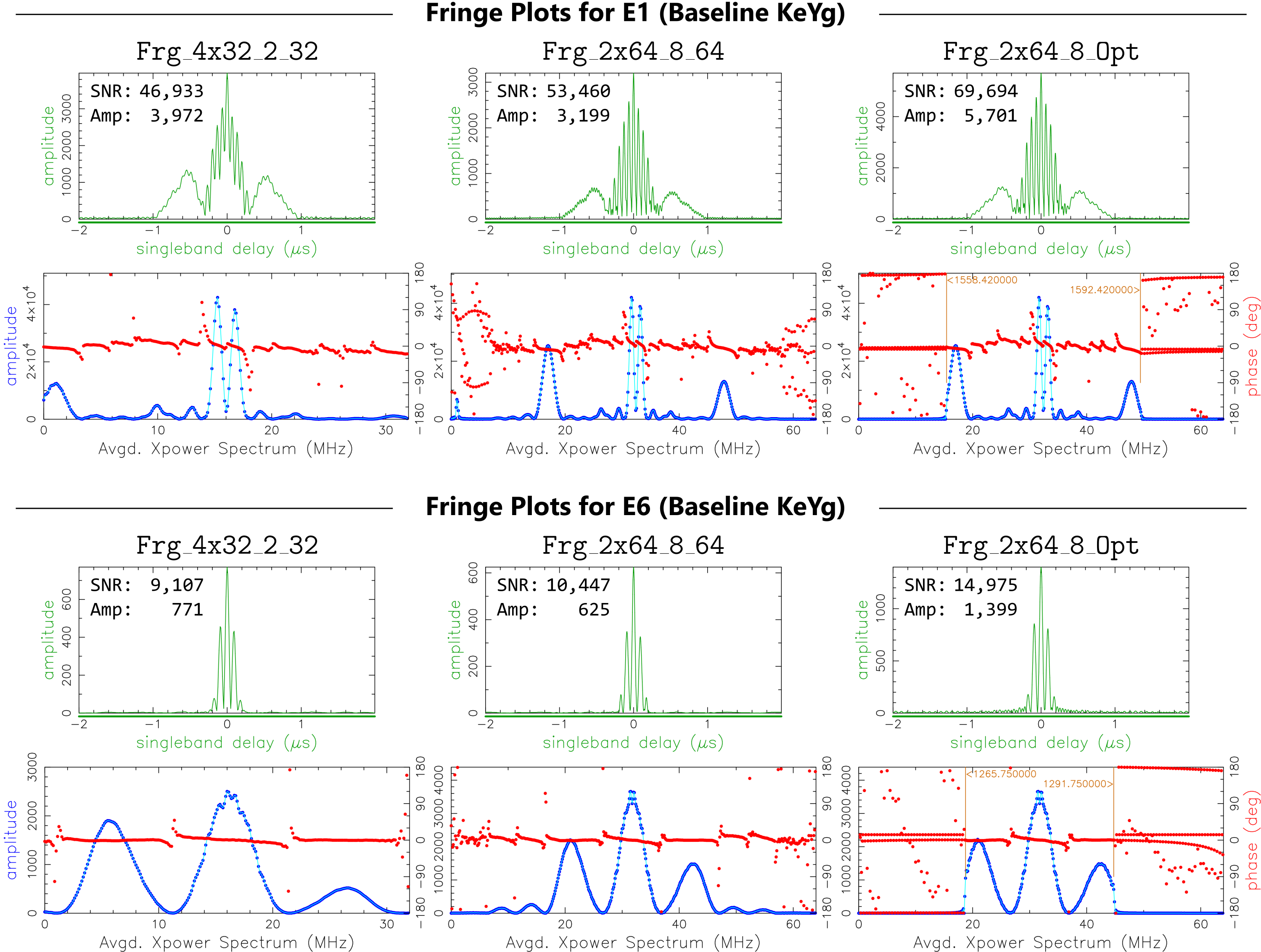}
	\caption{Fringe plots for single-bands in the E1 (top) and E6 (bottom) frequencies on the Katherine-Yarragadee baseline for the XX polarisation product showing the delay amplitudes as well as the amplitude and residual phase of the average cross-correlation spectra for the fringe modes \texttt{Frg\_4x32\_2\_32} (left), \texttt{Frg\_2x64\_8\_64} (centre) and \texttt{Frg\_2x64\_8\_Opt} (right). The observations are 5-min scans to the Galileo satellite with Pseudo-Random Noise (PRN) identifier E10.}
	\label{fig:fourfit_example}
\end{figure*}

Figure~\ref{fig:fourfit_example} shows typical fringe plots for the E1 and E6 bands of observations to Galileo satellites for three different fringe modes.
As expected, the signal structures are far from flat and uniform. The delay resolution functions are dominated by the frequency spacing of the peaks in the cross-power spectra.
The small frequency separation of the two dominant peaks in E1 produces a wider delay-domain response, while the greater separation among the three peaks in E6 results in a more compact response.
The spacing of the individual signal components creates a comb-like structure in the delay domain with several peaks. This is what would be called ``sub-ambiguities" in typical multi-band delay fitting. For both E1 and E6, the centre peaks have a significantly larger amplitude than the adjacent sub-peaks.
While the delay amplitudes show a similar characteristic across all fringe modes, the individual delay spikes are more defined for the 8-bit fringe modes \texttt{Frg\_2x64\_8\_64} and \texttt{Frg\_2x64\_8\_Opt}, especially in the E1 frequency band.
The residual phase plots have discontinuities across the frequency range, likely caused by the modulation structure of the transmitted data.

In contrast to fringe mode \texttt{Frg\_4x32\_2\_32} using 32~MHz channels, the ends of the channel bandpass of fringe mode \texttt{Frg\_2x64\_8\_64} with 64~MHz channel widths show no signal and do not contribute strongly towards the delay estimation. Fringe mode \texttt{Frg\_2x64\_8\_Opt} uses tuned frequency ranges to optimise the bandpass with respect to SNR. By disregarding the channel ends for fringe fitting the SNR and delay amplitude increase significantly.
It is evident from the figure, that the SNRs in the E1 frequency band are about five times as large as the ones in the E6 frequency bands on the Katherine-Yarragadee baseline.

\subsection{Post-processing and analysis}

After fringe fitting was completed, the single-band total delays in the E1 and E6 frequency bands, $\tau_{E1}$ and $\tau_{E6}$, were extracted. Among other effects, the observables are affected by the dispersive time delay caused by the ionosphere. To derive the ionosphere-free total delay $\tau_{\text{if}}$, an ionosphere-free linear combination, as, for example, described in \citet{Alizadeh2013atm}, was applied as

\begin{align}
	\tau_{\text{if}} &= c_1 \cdot \tau_{E1} - c_2 \cdot \tau_{E6} \hspace{1.5em}\text{with}\\
	c_1 &= \dfrac{f_{E1}^2}{f_{E1}^2 - f_{E6}^2} = 2.9312 \hspace{1.5em}\text{and}\\
	c_2 &= \dfrac{f_{E6}^2}{f_{E1}^2 - f_{E6}^2} = 1.9312,
\end{align}

where $f_{E1}$, the centre frequency of E1, is 1575.42~MHz, and $f_{E6}$, the centre frequency of E6, is 1278.75~MHz. The frequency band E5b (channel~4) of recording mode \texttt{Rec\_4x32\_2} is not used for analysis due to weak fringe detections and SNRs.

\begin{table*}
	\centering
	\begin{tabular*}{\linewidth}{@{\extracolsep{\fill}} c|llllrll}
		\toprule
		Type & Label & Code & Date & Start [UTC] & Dur [h] & Rec Mode & Antennas\\ \midrule
		\multirow{2}{*}{A} & A1 & as263a & 20/09 & 10:30 & $\sim$0.4 & \texttt{Rec\_2x64\_8} & HbKeYg\\
		& A2 & as263b & 20/09 & 12:00 & $\sim$0.4 & \texttt{Rec\_4x32\_2} & HbKeYg\\ \midrule
		\multirow{4}{*}{B} & B1 & as207a & 26/07 & 01:00 & 24        & \texttt{Rec\_4x32\_2} & HbKeYg\\
		& B2 & as298a & 25/10 & 11:00 & 24        & \texttt{Rec\_2x64\_8} & HbKeYg\\
		& B3 & as304a & 31/10 & 12:00 & 24        & \texttt{Rec\_2x64\_8} & HbKeYg\\
		& B4 & as305a & 01/11 & 12:00 & 24        & \texttt{Rec\_2x64\_8} & HbKeYg\\
		\bottomrule
	\end{tabular*}
	\caption{Overview of the VLBI observing sessions conducted in the scope of this study. Given are the session type, label, code, start time in UTC, duration in hours, recording mode and participating antennas. All sessions were conducted in 2025.}
	\label{tab:session_overview}
\end{table*}

The observables were analysed using the VLBI software package VieVS (Version 4.0). Besides the total delays, the satellite orbits were provided in the form of SP3 files. As described in Sect.~\ref{sec:02_interferometric_models}, VieVS has the Klioner near-field delay model \citep{Klioner1991agu} implemented to evaluate observations to near-field targets. The model is based on a light time solution of travel times. VieVS is capable to process both satellite and natural geodetic sources in the same session for the estimation of the model parameters. In this work, we only process satellite observations.

\section{Overview of experiments}\label{sec:03_overview_experiments}

Table~\ref{tab:session_overview} provides an overview of the VLBI experiments conducted in the scope of this study.
The six-letter session codes were used internally for the implementation and processing of the experiments. However, for convenience, in the following the sessions are referred to by their labels.
Two types of sessions were conducted. Type A sessions are short experiments with a duration of about 0.4~h, designed to test different aspects inherent to observations of satellite sources.
On the other hand, type B sessions are 24-h experiments, conducted with the aim to perform an analysis and estimate geodetic parameters with best possible accuracy.

\begin{table}[b!]
	\centering
	\begin{tabular*}{\columnwidth}{@{\extracolsep{\fill}} c|rrcclr}
		\toprule
		Label & \# & Start & Src & Dur & Rec Mode & Pnt \\ \midrule
		\multirow{4}{*}{A1} & 1 & 10:30 & \multirow{4}{*}{E10} & \multirow{4}{*}{300} & \multirow{4}{*}{\texttt{Rec\_2x64\_8}} & 30 \\
		& 2 & 10:36 & & & & 20 \\
		& 3 & 10:42 & & & & 10 \\
		& 4 & 10:48 & & & & 5 \\ \midrule
		\multirow{4}{*}{A2} & 1 & 12:00 & \multirow{4}{*}{E10} & \multirow{4}{*}{300} & \multirow{4}{*}{\texttt{Rec\_4x32\_2}} & 30 \\
		& 2 & 12:06 & & & & 20 \\
		& 3 & 12:12 & & & & 10 \\
		& 4 & 12:18 & & & & 5 \\
		\bottomrule
	\end{tabular*}
	\caption{Schedule summaries of the experiments A1 and A2 providing information about the experiment label, scan numbers (\#), scan start times in UTC time, satellite source name, scan duration in seconds, recording mode and pointing update intervals in seconds.}
	\label{tab:type_a_sessions}
\end{table}

\begin{figure}[b!]
\centering
\includegraphics[width=\linewidth]{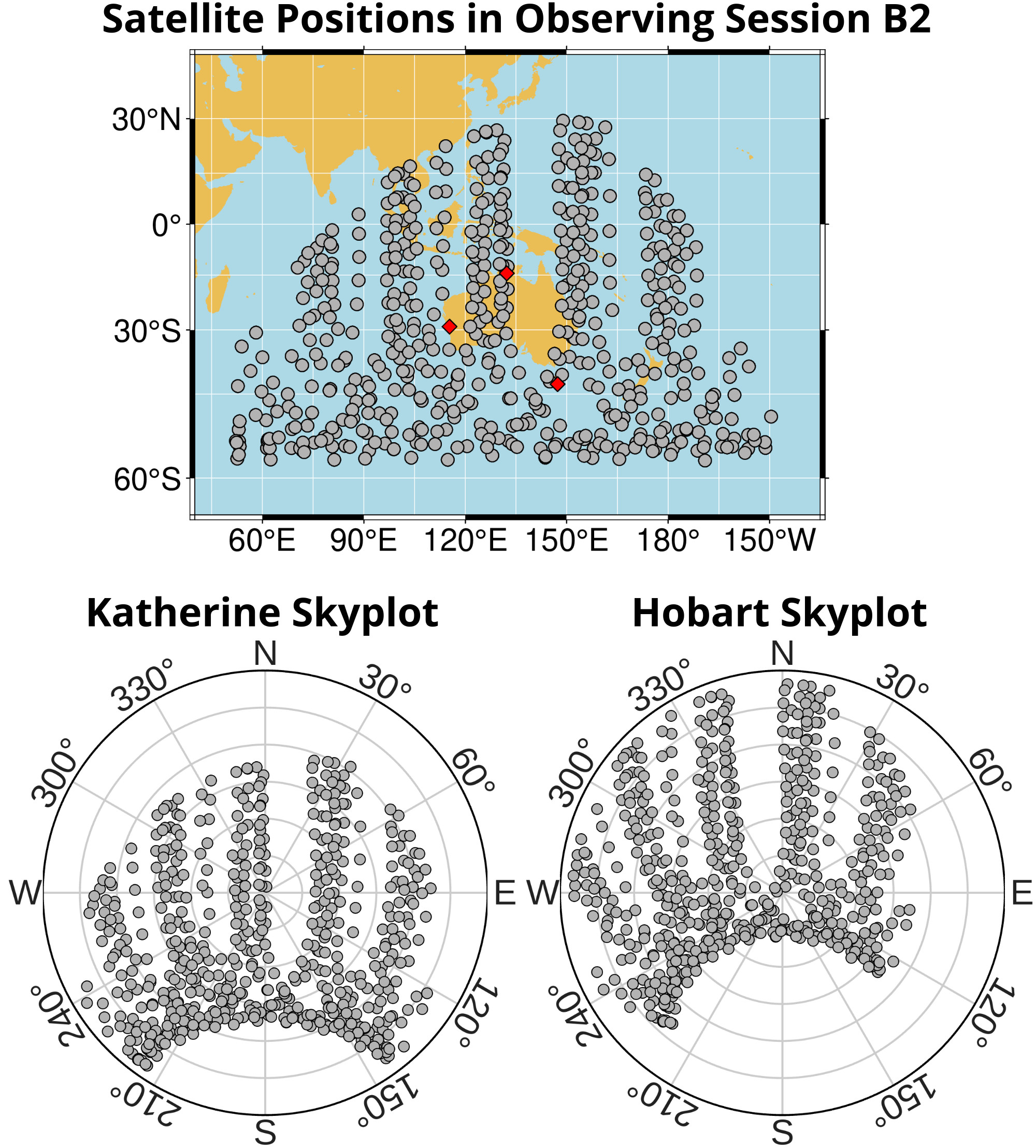}
\caption{Observing geometry of session B2. The positions of the Galileo satellites at the time of each scan are visualised as latitudes and longitudes on a world map projection (top) along with the topocentric view angles in the skyplots for the Katherine and Hobart antennas (bottom).}
\label{fig:experiment_as298a}
\end{figure}

The sessions A1 and A2 were executed consecutively with about one hour in between to allow for a change of the recording setup. Both sessions use the Galileo satellite with PRN identifier E10 (GSAT0224) as the only source. The schedule summaries for the type~A experiments are given in Table~\ref{tab:type_a_sessions}. The sessions were scheduled manually and consist of only four scans each. The scans have a duration of 300~s and only differ in the time intervals with which the pointing of the antenna in the direction of the satellite is updated. The intervals were set to 30, 20, 10 and 5~s, respectively. A comparison of the pointing update intervals can be used to investigate the impact of the stepwise tracking approach on the total delay.
Session A1 was observed in the 8-bit recording mode \texttt{Rec\_2x64\_8}, while the session A2 used the 2-bit recording mode \texttt{Rec\_4x32\_2}. The difference in recording modes allows for an investigation of the impact of the sample bit depth on the VLBI observable.
Additionally, the long scan durations of five minutes in the type~A experiments allow for a comparison and evaluation of the interferometric input model to the correlator with regard to their consistency and agreement with the data.

Type~B sessions consist of four full-scale 24-h experiments.
Session B1 was observed in the 2-bit recording mode \texttt{Rec\_4x32\_2} and the scan lengths were set to 30~s, resulting in about 960~scans.
The session B2--B4 were recorded in the 8-bit recording mode \texttt{Rec\_2x64\_8}. For these sessions, the scan lengths were set to 120~s, resulting in about 500~scans per session.

Figure~\ref{fig:experiment_as298a} visualises the positions of the Galileo satellites at the time of each scan as latitudes and longitudes on a world map projection along with the topocentric view angles in the skyplots for the Katherine and Hobart antennas for observing session B2. The skyplot for Hobart reveals a lack of satellite coverage in the east and south. The regional nature of the network limits the common visibility of satellites for all stations, preventing the scheduling of scans at low elevations in some directions. For Hobart, this leads to the absence of any scans lower than 30$^{\circ}$ elevation between 60$^{\circ}$ and 210$^{\circ}$ azimuth. Additionally, due to the limited orbit inclination of Galileo satellites of about 56$^{\circ}$, no satellite coverage is available in the high polar latitudes, resulting in a ``polar hole". This can degrade the performance of the estimation of station positions and is, due to its southern location, most pronounced in the skyplot of the Hobart antenna. The polar hole is less prominent in the Katherine skyplot due to its near equatorial location. Between about 300$^{\circ}$ and 30$^{\circ}$ azimuth, no scans were performed at an elevation below 30$^{\circ}$.

\section{Results}\label{sec:04_results}

\subsection{Comparison of recording and fringe modes}\label{sec:04_comp_rec_frg}

\begin{figure}[b!]
	\centering
	\includegraphics[width=1\linewidth]{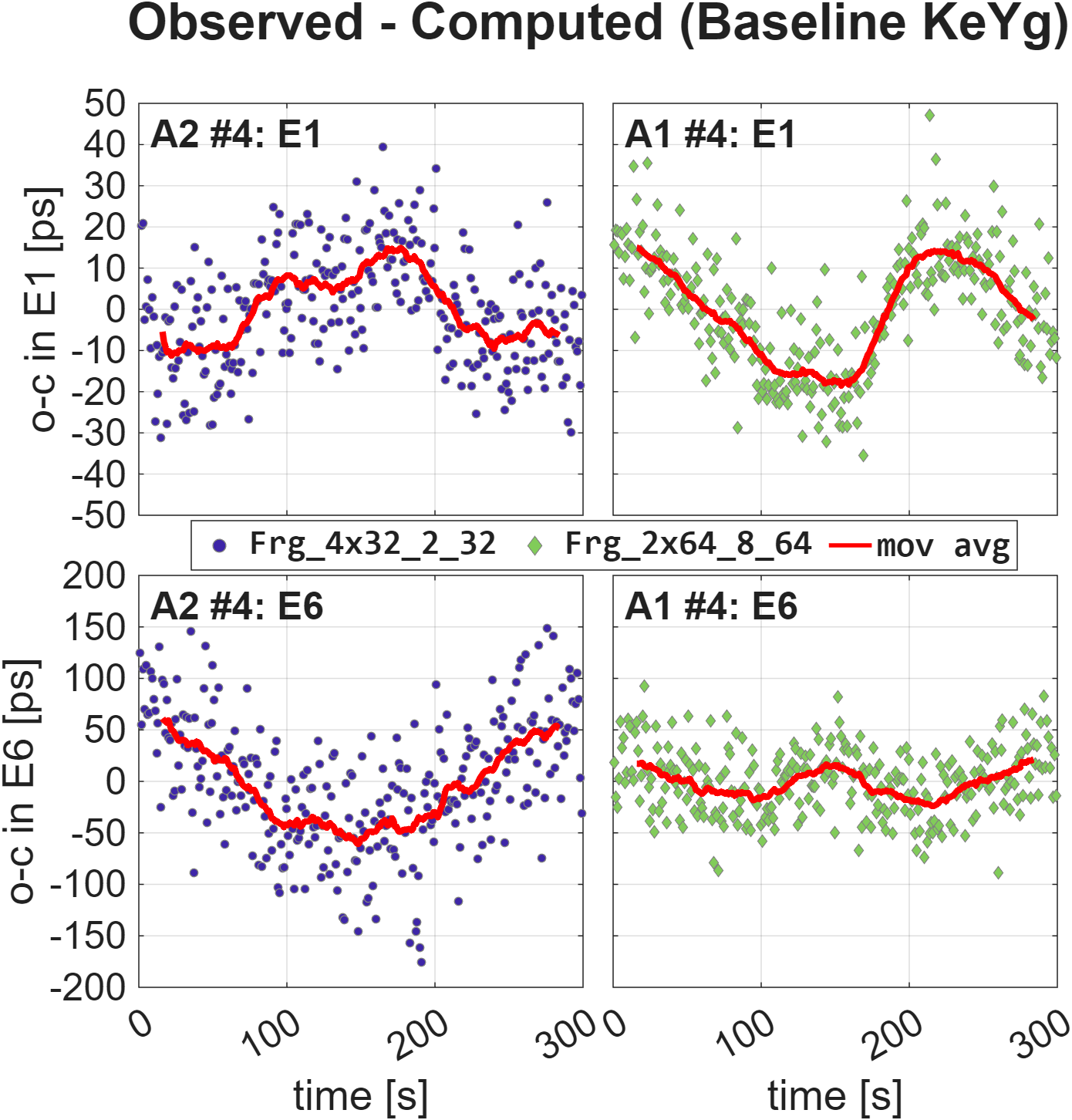}
	\caption{Observed-computed values of 300 1-s observations of scans~\#4 of the experiments A2 using fringe mode \texttt{Frg\_4x32\_2\_32} (left) and A1 using fringe mode \texttt{Frg\_2x64\_8\_64} (right) on the baseline Katherine-Yarragadee for the frequency bands E1 (top) and E6 (bottom). The red lines represent a moving average of the values using a window size of 30~s.}
	\label{fig:moving_average_example}
\end{figure}

\begin{figure*}
\centering
\includegraphics[width=1\linewidth]{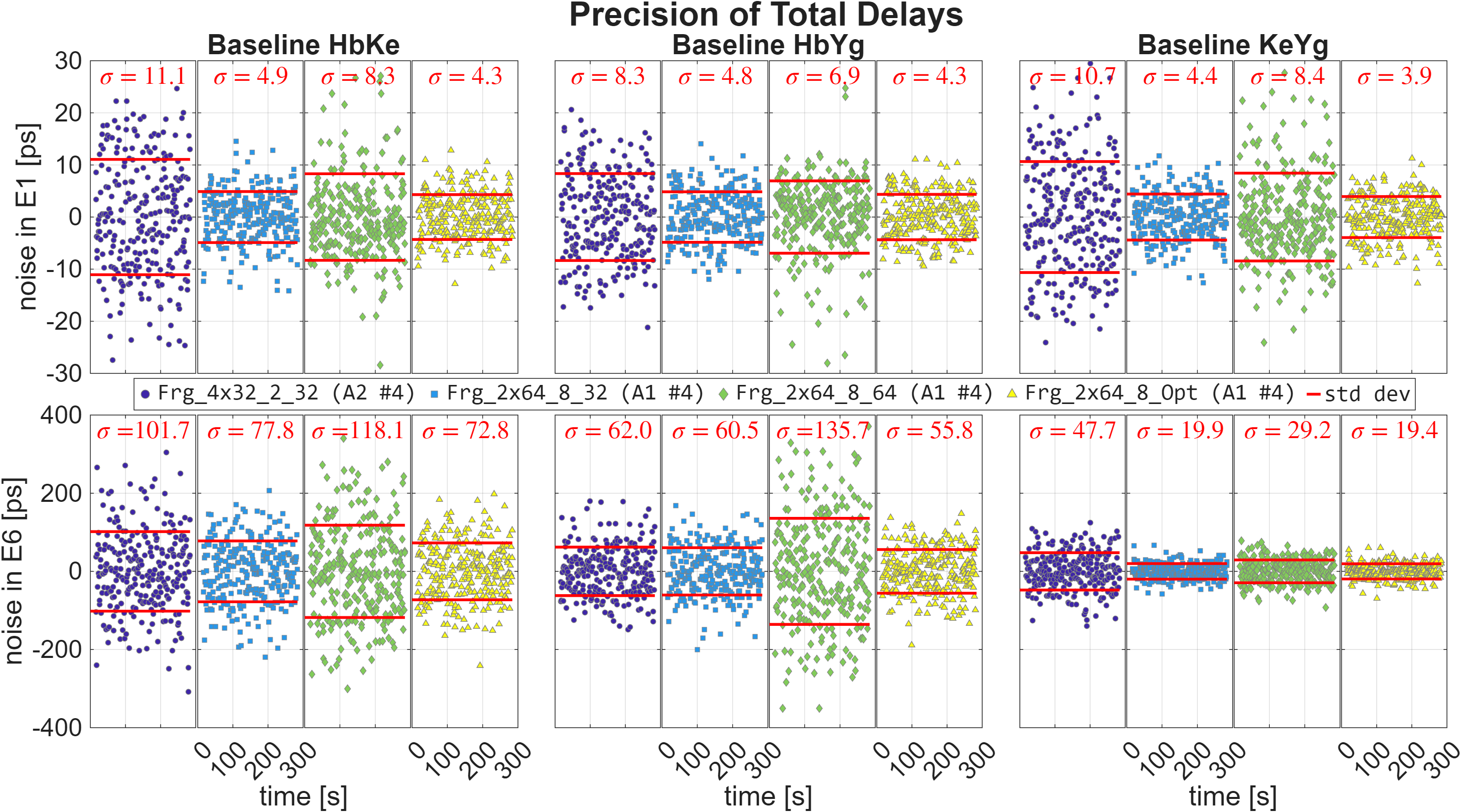}
\caption{Empirically determined precision of total delays for four different processing modes in the frequency bands E1 and E6 for a 5-min scan. The 2-bit recording of scan \#4 of session A2 is processed with fringe mode \texttt{Frg\_4x32\_2\_32}. The 8-bit recording of scan \#4 of session A1 is processed with the fringe modes \texttt{Frg\_2x64\_8\_32}, \texttt{Frg\_2x64\_8\_64} and \texttt{Frg\_2x64\_8\_Opt}. The red lines represent the standard deviations which are also quantified at the top of each subplot.}
\label{fig:precision_comparison}
\end{figure*}

The main difference between the A1 and A2 experiments is the employed recording mode. Assuming that the difference in the experiment start times of 1.5~hours has only a small effect on the results, the two experiments can be used to directly compare the 2-bit with the 8-bit recording mode and evaluate the impact of the introduced fringe modes. The comparison is performed based on the measured delays. As the absolute delays differ due to a change of the observing geometry between the two sessions, in the following, the total delays are compared with respect to their precision. For this, the fourth scans of the A1 and A2 experiments were divided into 300 1-s scans. The residuals (observed-minus-computed values) for both scans are shown for one polarisation product of the E1 and E6 frequency bands on the Katherine-Yarragadee baseline in Fig.~\ref{fig:moving_average_example}. The computed values were determined with VieVS. A constant rate and offset were removed from the residuals, and a moving average with a window size of 30~s was determined. Besides noise, the residuals show a systematic variation of the residual delay, emphasised in the moving average. As the signals in the E1 frequency band are not similar or equivalent to the signals in their respective E6 frequency bands, it is assumed not to be caused by tropospheric or ionospheric delays. Instead, they could originate from the satellite or be inadvertently introduced in the L-band signal chain. These signals are discussed in Sect.~\ref{sec:05_dicussion}.

The noise is determined by removing the moving average from the residuals. Assuming that the moving average represents the majority of the underlying signal, the remaining noise is used to quantify the precision of the delays. Figure~\ref{fig:precision_comparison} compares the precision of the 2-bit and 8-bit recording mode as well as the fringe modes and presents their standard deviation. The precision for the fourth scan of session A2 is determined with the fringe mode \texttt{Frg\_4x32\_2\_32}. For the fourth scan of session A1, the precision is determined for the fringe modes \texttt{Frg\_2x64\_8\_32}, \texttt{Frg\_2x64\_8\_64} and \texttt{Frg\_2x64\_8\_Opt}. Overall, the noise in the E1 frequency band ranges from a minimum of 3.9~ps to a maximum of 11.1~ps. In contrast, the noise is significantly higher in the E6 frequency bands, ranging from 19.4~ps to 135.7~ps. This can be attributed, on the one hand, to the reduced gain of the 12-m antennas at lower frequencies in L-band and, on the other hand, to the lower transmitted signal power of the Galileo E6 signal compared to E1 at the satellite \citep[e.g.][]{Steigenberger2018jog}.

The result for the scans of the A2 and A1 sessions with fringe modes \texttt{Frg\_4x32\_2\_32} and \texttt{Frg\_2x64\_8\_32} use the same bandpass for fringe fitting, allowing a direct comparison of the 2-bit with the 8-bit recording mode using the same frequencies. In the E1 frequency band the noise in the 2-bit recordings is about twice as large as the noise in the 8-bit recordings. For example, on the Hobart-Katherine baseline they are 11.1~ps and 4.9~ps. For the E6 frequency band, the differences are not as significant. While the noise is more than twice as large in 2-bit in comparison to 8-bit on the Katherine-Yarragadee baseline, it is only a factor of 1.3 larger on the Hobart-Katherine baseline and of similar magnitude on the Hobart-Yarragadee baseline. 
These results suggest that the 8-bit recordings are more advantageous in E1 than in E6. This can be attributed to the greater relative strength of the E1 signal. The increased dynamic range of the greater bit-depth is more effective at reducing saturation effects.

It is evident that across all three baselines and both frequency bands the fringe mode \texttt{Frg\_2x64\_8\_64} leads to the worst precision among the tested fringe modes for the A1 experiment with 8-bit data. The mode includes the entire recorded channel width of 64~MHz for fringe fitting. As shown in Fig.~\ref{fig:fourfit_example}, the outer edges of the channels contain no or only weak navigation signals. By reducing the passband from 64~MHz (\texttt{Frg\_2x64\_8\_64}) to 32~MHz (\texttt{Frg\_2x64\_8\_32}) the fringe fitting includes the strongest signals in the cross correlation spectra while disregarding the frequencies with smaller power. This leads to an increase in SNR and, consequently, an improved precision. For example, the precision improves from 135.7~ps to 60.5~ps in the E6 frequency band on the Hobart-Yarragadee baseline. The precision further improves to 55.8~ps with the optimised fringe mode (\texttt{Frg\_2x64\_8\_Opt}), which uses an empirically tuned bandpass. Overall, the fringe mode \texttt{Frg\_2x64\_8\_Opt} generates the highest SNR values and best precisions for both frequencies and all baselines. Hence, in the following investigations, this mode is used for all investigations of 8-bit data.

The fact that a delay precision on the order of a few picoseconds in E1 and well below 100~ps in E6 can be achieved at 1-s coherent integration time using a single channel for correlation underlines the received signal strength of the Galileo satellites in contrast to the faint signals of natural geodetic radio sources in standard geodetic VLBI experiments. According to \cite{Whitney1974phd} the theoretical precision of the group delay is determined by the effective bandwidth $\Delta v_{RMS}$ of the receiving system and the SNR as
\begin{align}
	\sigma_{\tau} = \dfrac{1}{2\pi \cdot SNR \cdot \Delta v_{RMS}}.\label{eq:precision}
\end{align}
The SNR, given by
\begin{align}
	SNR \equiv \rho_0 \sqrt{2 \cdot \Delta v_{RMS} \cdot T}, \label{eq:snr}
\end{align}
is equivalent to the product of the normalised correlation amplitude $\rho_0$, the spanned bandwidth $\Delta v_{RMS}$ and the effective integration time $T$. With the bandwidth restricted by the transmitted narrow-band signal of the Galileo satellites, the SNR is the varying factor that limits the precision.

\begin{table}[t!]
	\begin{tabular*}{\columnwidth}{@{\extracolsep{\fill}} rrrr}
		\toprule
		$T$ [s] &   $SNR_T$ & $R_{I} = \dfrac{SNR_T}{SNR_{T=1}}$ & $R_{II} = \sqrt{\dfrac{T}{T_{T=1}}}$ \\ \midrule
		1 &   $5\,082.5$ &  1.00 &  1.00 \\
		2 &   $7\,192.7$ &  1.41 &  1.41 \\
		5 &  $11\,360.1$ &  2.24 &  2.24 \\
		10 &  $16\,088.3$ &  3.17 &  3.16 \\
		30 &  $27\,859.9$ &  5.48 &  5.48 \\
		60 &  $39\,362.8$ &  7.74 &  7.75 \\
		120 &  $55\,723.4$ & 10.96 & 10.95 \\
		180 &  $67\,954.4$ & 13.37 & 13.42 \\
		298 &  $87\,263.4$ & 17.17 & 17.26 \\
		\bottomrule
	\end{tabular*}
	\caption{Overview of the increase in SNR with respect to the integration time of observations to Galileo satellites in the E1 frequency band on the Hobart-Katherine baseline. Given are the integration times $T$ in seconds and the respective SNR along with the ratio $R_{I}$ of the measured increase in SNR with respect to the SNR at an integration time of 1~s and the theoretical ratio $R_{II}$ of the expected increase in SNR based on the integration times.}
	\label{tab:int_snr}
\end{table}

Using the identical experimental setup, the SNR can be increased with longer integration times according to Eq.~\ref{eq:snr}. To test whether this holds not only for the noise-like electromagnetic radiation of natural geodetic radio sources but also for the deterministic signals of GNSS satellites, the increase in SNR with respect to the coherent integration time is analysed. Table~\ref{tab:int_snr} shows the integration time $T$ in seconds along with its respective measured SNR ($SNR_T$) as provided through \texttt{fourfit}.
Furthermore, the ratio $R_{I}$ of the measured increase in SNR with respect to the SNR at an integration time of 1~s ($SNR_{T=1}$) and the theoretical ratio $R_{II}$ of the expected increase in SNR based on the integration times are provided. The values were derived from the fourth scan of session A1 on the Hobart-Katherine baseline in the E1 frequency band by limiting the integration time in the fringe fitting process. The SNR increases from $5\,082.5$ at 1~s to $87\,263.4$ at 298~s. Of the scheduled 300~s only 298.7~s were recorded due to a delay at the start of recording caused by the processing time of the tracking command in the Field system. The empirically derived values of $R_{I}$ agree well with the theoretical ones of $R_{II}$, indicating that Eq.~\ref{eq:snr} holds true for the observations of Galileo satellites. 
According to Eq.~\ref{eq:precision}, satellite scans with integration times of 30 and 120~s should therefore achieve precisions that are more than a factor of 5 and 10 higher, respectively, than those shown in Fig.~\ref{fig:precision_comparison}.

\subsection{Polarisation products}

\begin{figure}[t!]
	\centering
	\includegraphics[width=\linewidth]{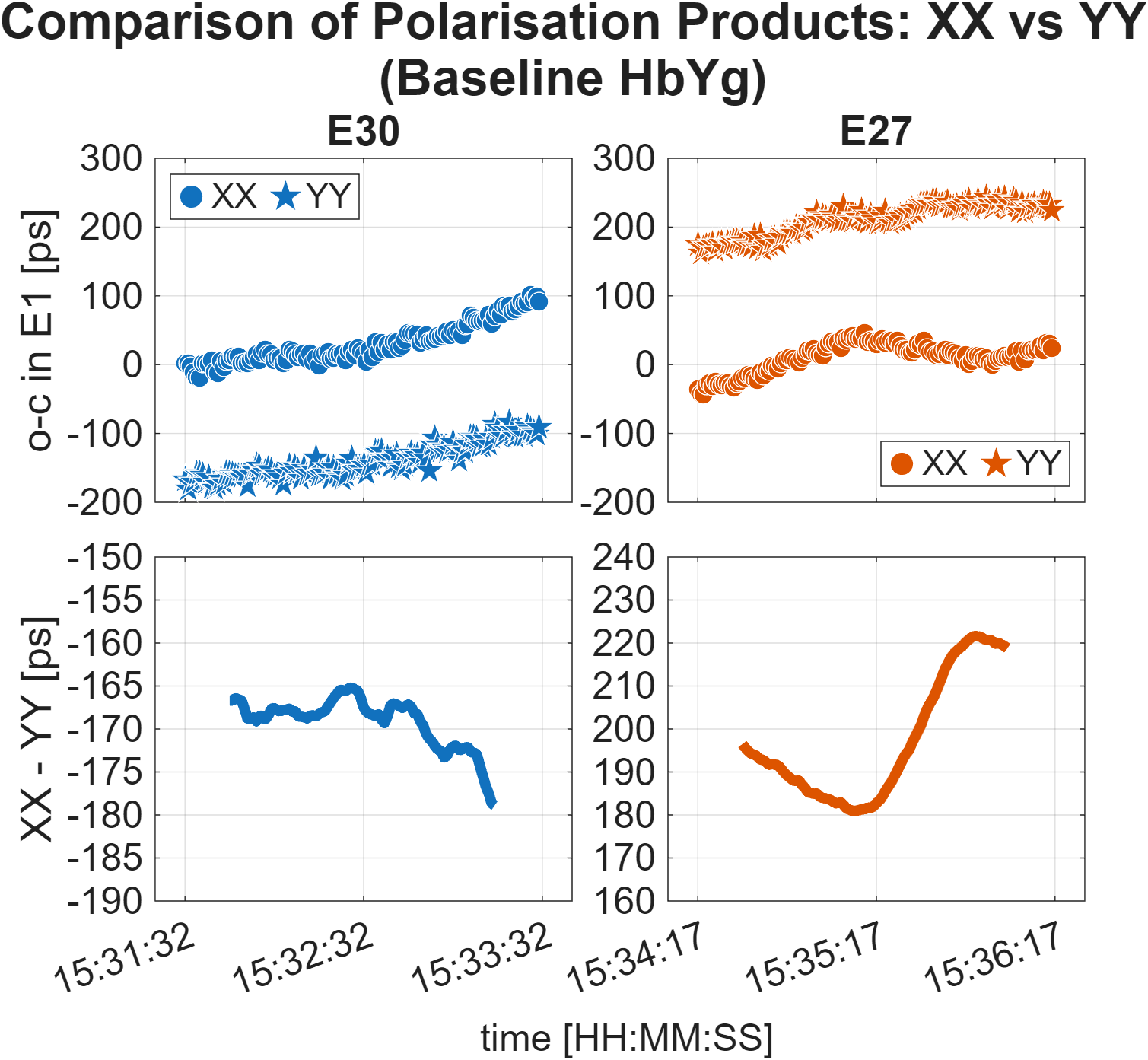}
	\caption{Comparison of the polarisation products XX and YY on two typical scans from the B3 session to the E30 (left) and E27 satellites (right) on the Hobart-Yarragadee baseline in the E1 frequency band. Shown are the residuals for the polarisation products in XX and YY (top). Furthermore, the differences between the residuals are visualised (bottom). To isolate the underlying trend from the noise, the differences were smoothed using a moving average filter with a 30-s window.}
	\label{fig:pol_comp_ex}
\end{figure}

The data was recorded in two linear polarisations at each station. By correlating the individual data streams for each baseline four individual polarisation products were formed, namely XX, YY, XY and YX. In the following, the agreement and consistency of these products with respect to each other are investigated. A period of about 40~min of data from the B3 session is considered in more detail. For this period, the data were divided into 1-s scans and residuals were formed for the XX and YY polarisation products.
Figure~\ref{fig:pol_comp_ex} demonstrated the processing based on two scans.
The figure compares the XX and YY polarisation products on two typical scans from the B3 session to the E30 and E27 satellites on the Hobart-Yarragadee baseline.
The upper panels in the figure display the residuals for both polarisation products.
The lower panels present the differences of the residuals in XX and YY. To isolate the underlying trend from the noise, the differences were smoothed using a moving average filter with a 30-s window.
For the scan to the E30 satellite, the differences are not constant over time but differ from about $-165$~ps to $-180$~ps. A variation of 15~ps is the equivalent of 4.5~mm. For the scan to the E27 satellite, the differences are positive and vary significantly more from about 180~ps to 220~ps, a variation equivalent to 12~mm.

\begin{figure*}
	\centering
	\includegraphics[width=1\linewidth]{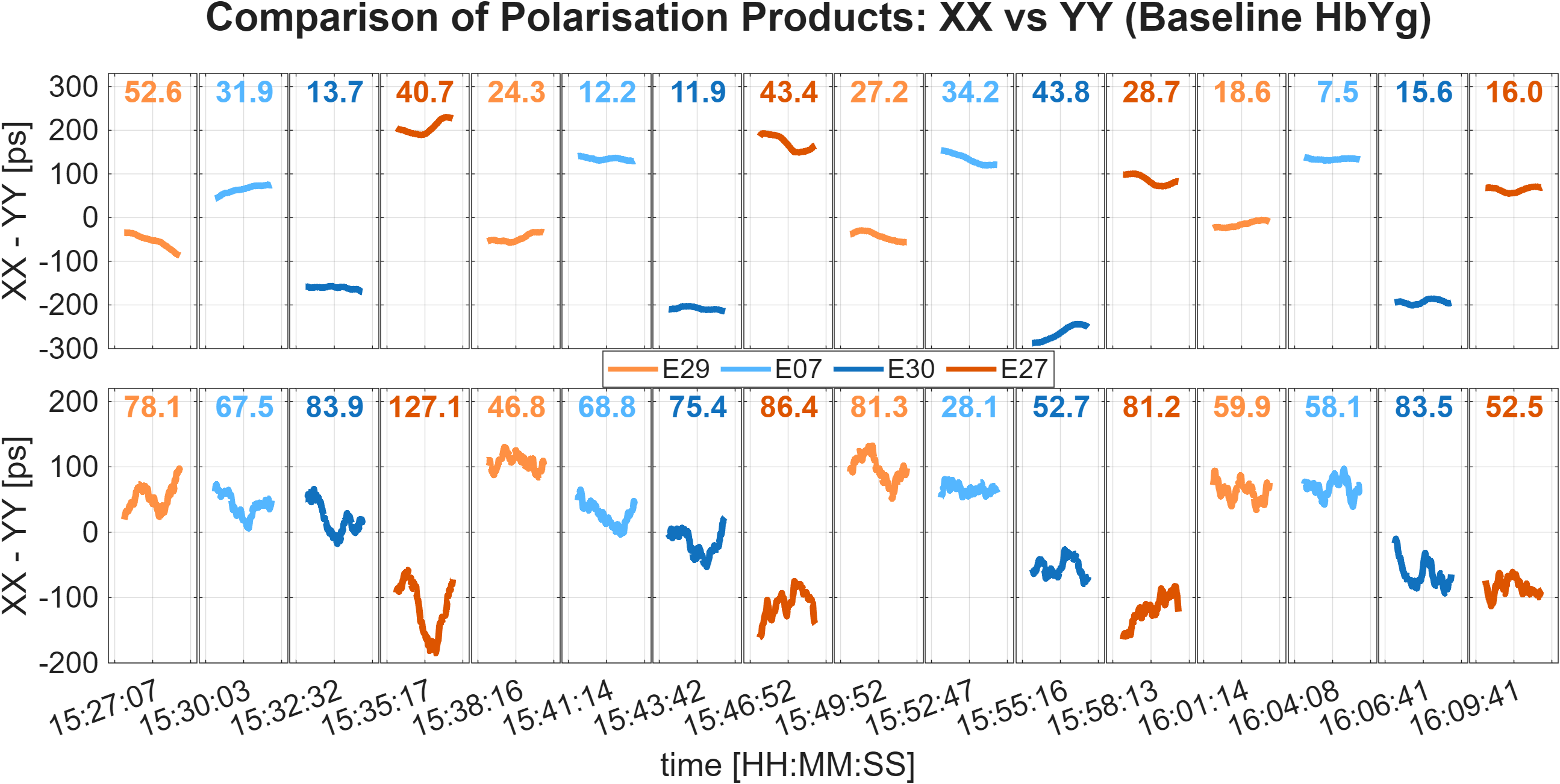}
	\caption{Differences of the total delays of the XX and YY polarisation products from a 40-min period of the B3 session on the Hobart-Yarragadee baseline in the E1 and E6 frequency bands. The maximal variation representing the difference between the minimum and maximum difference is provided at the top of each individual subplot.}
	\label{fig:pol_diff_xx_yy}
\end{figure*}

Figure~\ref{fig:pol_diff_xx_yy} visualises the difference from all scans within the 40-min period in the E1 and E6 frequency bands. This particular period was selected as only four satellites were visible and, hence, each satellite was observed frequently. A constant offset was removed between the total delays of the XX and YY products before building the differences, shifting them around zero.
For both frequency bands, the differences show satellite-dependent systematic effects.
For example, in E1 the differences for the satellite E30 are all negative between $-300$~ps and $-150$~ps, while the differences for the satellite E27 are all positive between 50~ps and 220~ps.
This effect can be observed for the entire selected period. Assuming it is not caused by the satellite's transmit antenna, which should have identical characteristics and performance across all Galileo satellites according to ESA, it indicates that it could be caused by the geometric configuration of the satellite and the observing antennas.
For the inspected period, the differences vary from about $-$300 to 230~ps in E1 and from about $-$200 to 140 in E6.

The differences do not behave consistently between the E1 and E6 frequency bands. For example, while the differences for E27 are positive in E1, they are negative in E6. And while the differences for E29 are negative for E1, they are positive in E6. This discrepancy between the two polarisation products indicates a frequency dependent effect.

Similar to the example scans in Fig.~\ref{fig:pol_comp_ex} the differences shown in Fig.~\ref{fig:pol_diff_xx_yy} do not remain constant but vary within the 2-min scan lengths. 
For individual scans, the differences are smooth in E1 while they are comparably noisy in E6. Furthermore, they vary significantly more within each scan in E6.
At the top of each subplot the variation of the differences is provided. They are derived from the minimum and maximum value of the differences.
For E1, the scan with the largest variation between the total delays in XX and YY is the scan at 15:27:07 to satellite E29 with about 52.6~ps, the equivalent of 15.8~mm.
For E6, the scan with the largest variation is the scan at 15:35:17 to satellite E27 with about 127.1~ps, the equivalent of 38.1~mm.
The variations in the differences could be caused by the underlying signals in the data as already mentioned in Sect.~\ref{sec:04_comp_rec_frg} and illustrated in Fig.~\ref{fig:moving_average_example}. They indicate that the signals differ on the order of several tens of picoseconds between the XX and YY polarisation product. In the E6 frequency band, the noise in the data and the amplitude of the underlying signal is larger compared to E1.

\subsection{Stepwise Tracking}

\begin{figure*}
	\centering
	\includegraphics[width=1\linewidth]{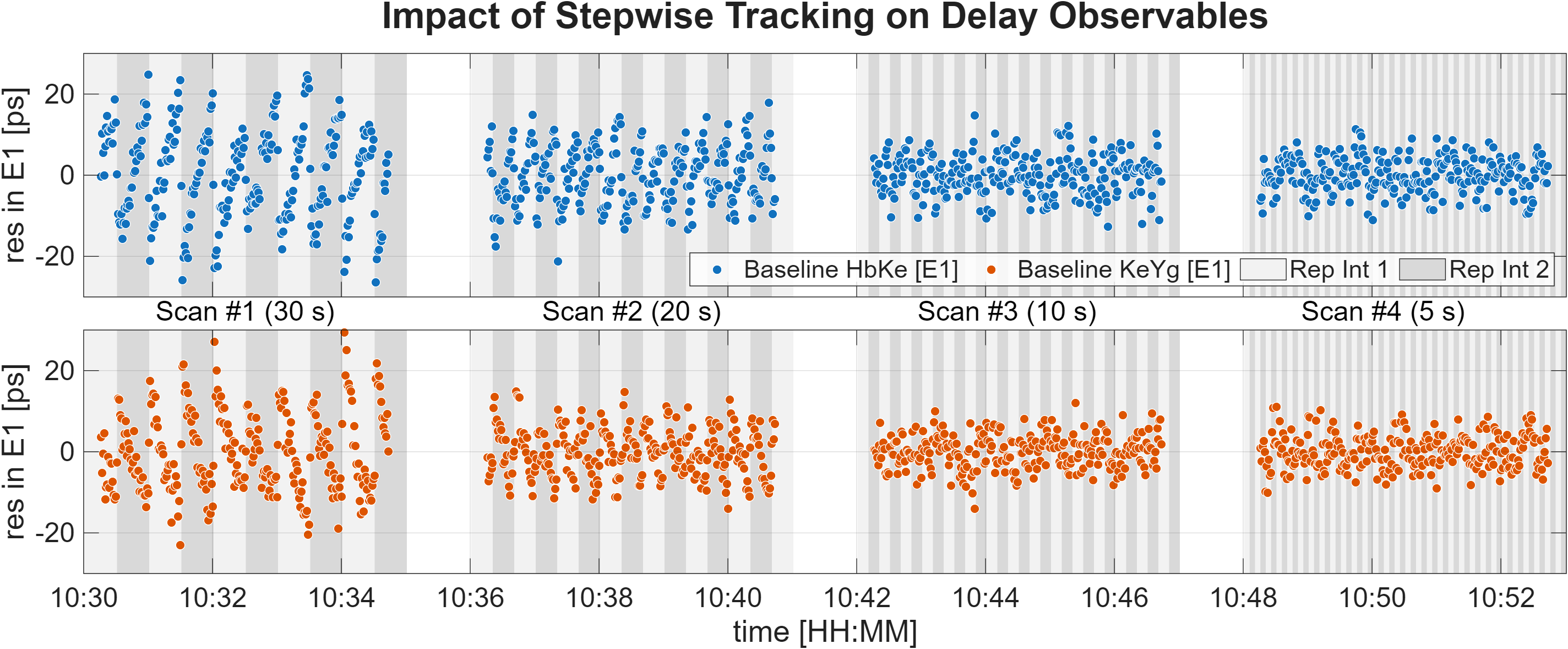}
	\caption{Impact of stepwise tracking on the delay observables for the scans 1--4 of the A1 experiment recorded in 8-bits on the Hobart-Katherine baseline (top) and Katherine-Yarragadee baseline (bottom) in the XX polarisation product. The residuals are derived by subtracting a moving average with a window size of 30~s from the total delays in the E1 frequency band.}
	\label{fig:oc_stepwise_tracking}
\end{figure*}

The utilised antennas in the experiments of this study are not yet able to track a satellite continuously. To investigate the impact of the employed stepwise tracking approach described in Sect.~\ref{sec:02_tracking} on the delay observables, the sessions A1 and A2 were designed to contain 5-min scans with tracking update intervals of 30, 20, 10 and 5~s, allowing a comparison of the different intervals.
As in the previous investigations, the scans are divided into 300 1-s scans and a moving average with a windows size of 30~s is subtracted from the residuals to remove the underlying signal in the data.
Figure~\ref{fig:oc_stepwise_tracking} visualises the residuals of the scans of session A1. The residuals are given for the Hobart-Katherine and Katherine-Yarragadee baselines in the E1 frequency band for one polarisation product. The figure also contains light and dark grey areas to illustrate the tracking update intervals.

The residuals of scan \#1 contain a notable systematic signal with a period equal to the tracking update interval of 30~s. The change in delay due to the stepwise tracking is noticeable as stripes that form within each interval.
While the antenna remains static for the 30-s interval, the satellite further moves along its orbit changing its angular position with respect to the antenna's beam.
This geometric change causes an increase and decrease of the incoming signal amplitude, leading to phase distortions. They manifest as a drift in the delay observable.
At the start of each interval the antenna receives the new position command, accelerates to the new position and decelerates when reaching the commanded position. This causes a jump in the observable, forming a striped pattern. For scan \#1, the stripes range from about $-25$ to $20$~ps for both presented baselines.

In the scan \#2, the striped pattern is still present although not as notable as for scan~\#1. Furthermore, in comparison, the amplitude in the stripes is reduced. Due to the shorter tracking update intervals, the satellite's angular distance to the antenna main beam is reduced at the beginning and end of the intervals, leading to smaller variations in the signal amplitude.
For the scans \#3 and \#4, the pattern is not noticeable anymore.
The variations in their residuals seem to be dominated by noise in the data. However, the fact that the stripes are clearly detectable for the 30-s and 20-s update intervals indicate that the intrinsic noise of the total delays may be smaller than the variation in the residuals shown for scans \#3 and \#4 with shorter update intervals. It it possible that a portion of the apparent noise in these scans is actually caused by the frequent acceleration and deceleration of the antenna, leading to the mentioned geometric effects that manifest as phase distortions.

\begin{figure}[b!]
	\centering
	\includegraphics[width=\linewidth]{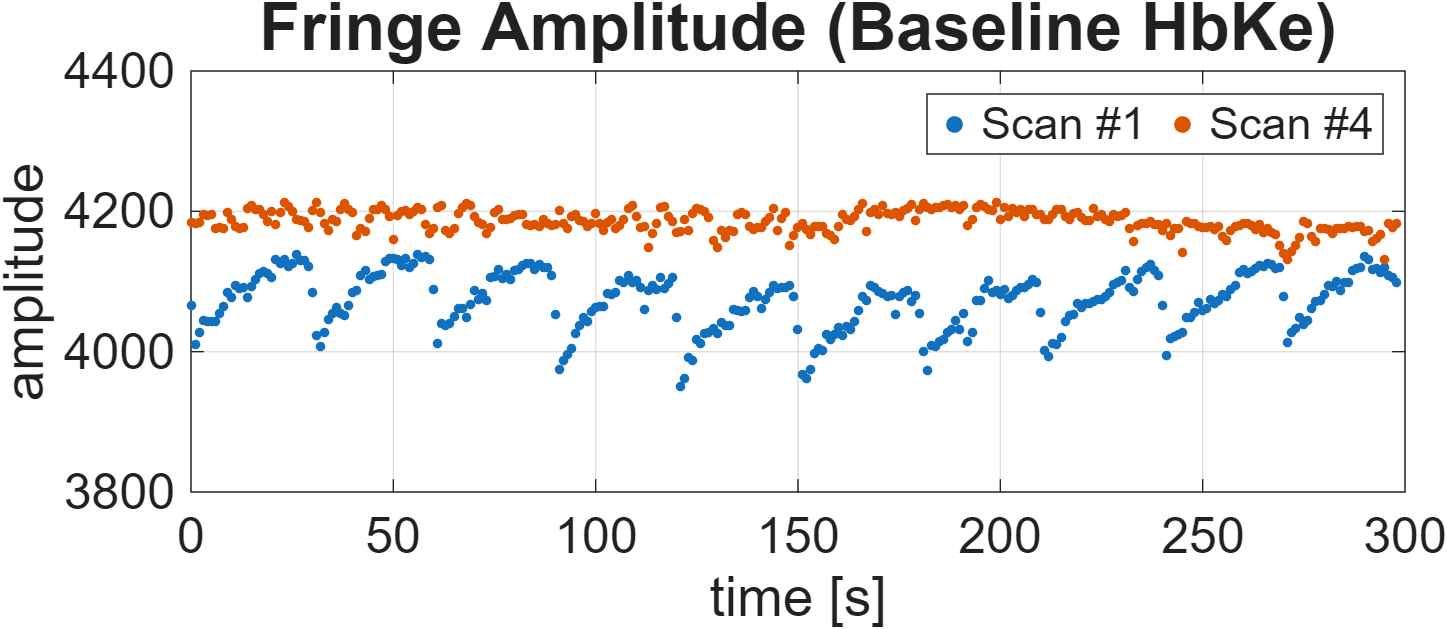}
	\caption{Comparison of the fringe amplitudes between scans \#1 (30-s tracking update interval) and \#4 (5-s tracking update interval) of session A1 recorded in 8-bit on the Hobart-Katherine baseline in the XX polarisation product.}
	\label{fig:modelscan1vsscan4}
\end{figure}

Figure~\ref{fig:modelscan1vsscan4} visualises the fringe amplitudes of scans \#1 and \#4 of session A1 on the Hobart-Katherine baseline. In scan \#1, a distinct periodic pattern is evident in the fringe amplitudes, reflecting the impact of the 30-s tracking update interval. While the fringe amplitudes in scan \#4 are more stable, they are not entirely smooth and display short-term variations. As previously mentioned, the frequent antenna repositioning commands of 5~s could cause these fringe amplitude fluctuations and consequently introduce phase distortions.

In the experimental setup, the chosen update interval and apparent motion are the main components to impact the amplitude in the effect of a striped pattern in the delay observables. Satellite positions at high elevations close to the antenna's zenith direction lead to fast angular rates in the topocentric view angles from the station. In the A1 session, the elevations during scan \#1 to \#4 changed from 63.7$^{\circ}$ to 72.2$^{\circ}$ at Hobart, 38.3$^{\circ}$ to 43.6$^{\circ}$ at Katherine and 56.1$^{\circ}$ to 57.7$^{\circ}$ at Yarragadee. As for all three antennas, the elevation increased during the 40-min experiment, the later scans of the experiment should be more affected by geometrical effects due to the faster apparent motion of the satellite.
However, because the elevation changes are less than 10$^{\circ}$, it is expected that the impact of the effect of an increasing apparent motion from scan \#1 to \#4 is small.

Figure~\ref{fig:oc_stepwise_tracking} only visualises the impact of the stepwise tracking on residuals in the E1 frequency. In the E6 frequency band, the striped patterns are not detectable. This is because the noise in the total delays is significantly larger than the effect itself. As investigated in Sect.~\ref{sec:04_comp_rec_frg}, the noise is at the level of a few picoseconds in E1, while it values up to several tens of picoseconds in E6.

As the stepwise tracking does not show a considerable impact on the delays in the scan \#3, an update interval of 10~s was chosen for the full-scale 24-h experiments B1--B4. Due to the high rate of positioning commands, an update interval of 5~s was ruled out as it was considered to put too much stress on the Field System, potentially risking it to become overloaded and unresponsive.

\subsection{Impact of near-field delay model}

Accurate interferometric input models are crucial for the correlator to align the data streams and retain coherence throughout the duration of the scan. In Sect.~\ref{sec:02_interferometric_models}, three methods were introduced to derive the a priori near-field delay models for the correlation of VLBI observations of Earth-orbiting satellites. In the following, the impact of the near-field delay model implementation on the results is investigated.

\begin{figure}[t!]
	\centering
	\includegraphics[width=\linewidth]{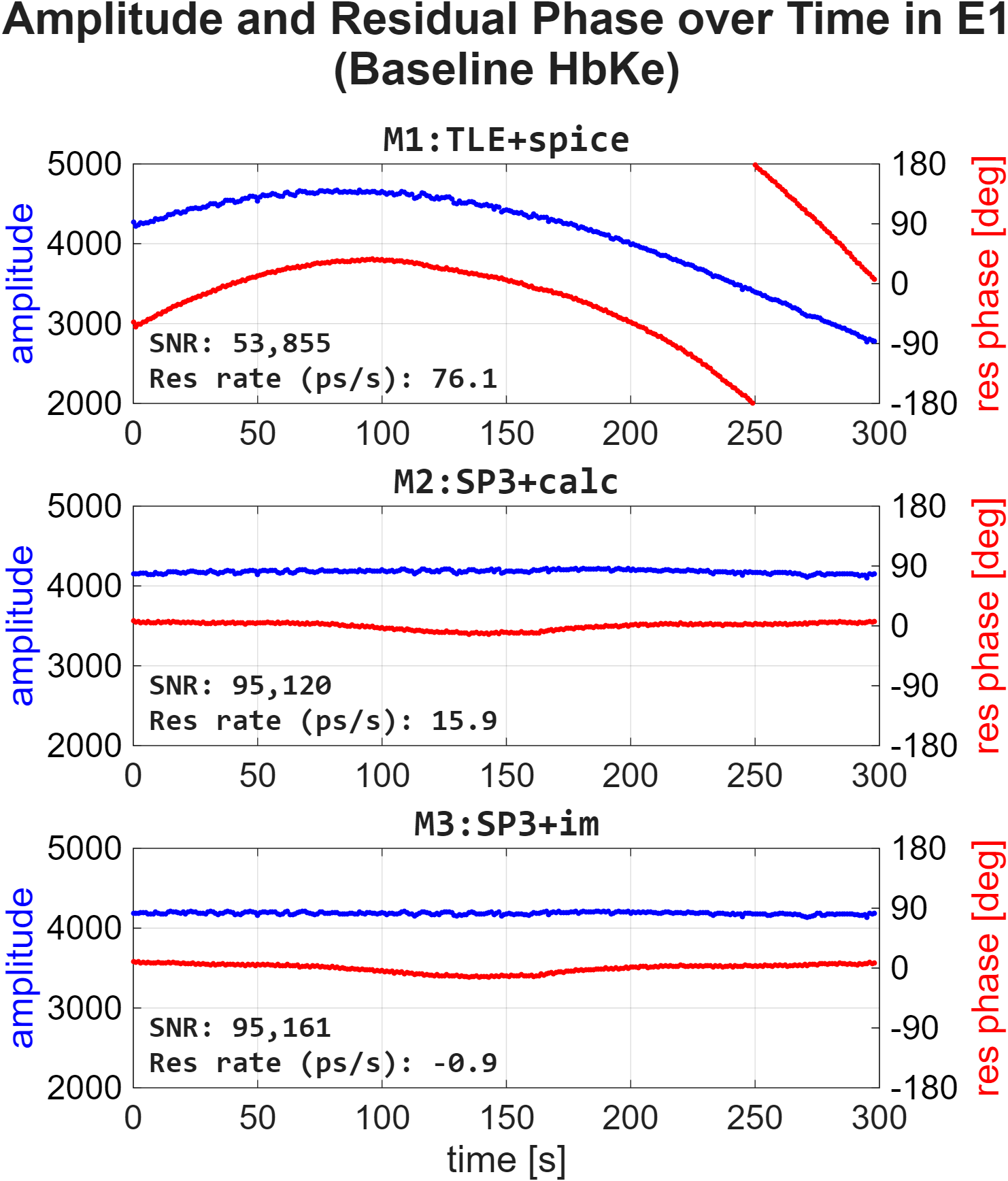}
	\caption{Fringe amplitude and residual phase over time of a single 5-min scan (scan~\#4 of experiment A1) in the frequency band E1 for the correlation with the interferometric delay model implementations \texttt{M1:TLE+spice} (top), \texttt{M2:SP3+calc} (middle) and \texttt{M3:SP3+im} (bottom).}
	\label{fig:delay_model_comparison}
\end{figure}

Figure~\ref{fig:delay_model_comparison} shows the amplitude and residual phase over time in the E1 frequency band for scan \#4 of the A1 experiment for the correlation with the interferometric delay model implementations \texttt{M1:TLE+spice}, \texttt{M2:SP3+calc} and \texttt{M3:SP3+im}. For the implementation based on TLEs, the fringe amplitude is not constant throughout the duration of the scan but decreases towards the end. The residual phase is not constant either and varies by more than 360$^{\circ}$. The SNR is $53,855$ and the residual phase rate equals 76.1~ps/s.
The large residual phase rate and the fringe amplitude drop indicate that there is a loss of coherence towards the end of the scan. It appears that either the delay or delay rate are larger than can be estimated from the frequency and time resolution of the correlated data when applying \texttt{M1:TLE+spice}.

In contrast, the delay model implementations based on SP3 files lead to significantly improved results. For both \texttt{M2:SP3+calc} and \texttt{M3:SP3+im}, the fringe amplitude remains mostly constant throughout the scan and the variations in the residual phase are minimal. Furthermore, the SNRs equal $95,120$ and $95,161$, an increase of about 80\% with respect to the TLE-based model. While \texttt{M2:SP3+calc} provides a residual phase rate of 15.9~ps/s, the value for \texttt{M3:SP3+im} is significantly smaller at $-0.9$~ps/s, demonstrating a good match of the delay model to the recorded data. Similar performance differences between the model implementations were found for the scans of the full-scale experiments B1--B4.

\begin{figure}[t!]
	\centering
	\includegraphics[width=\linewidth]{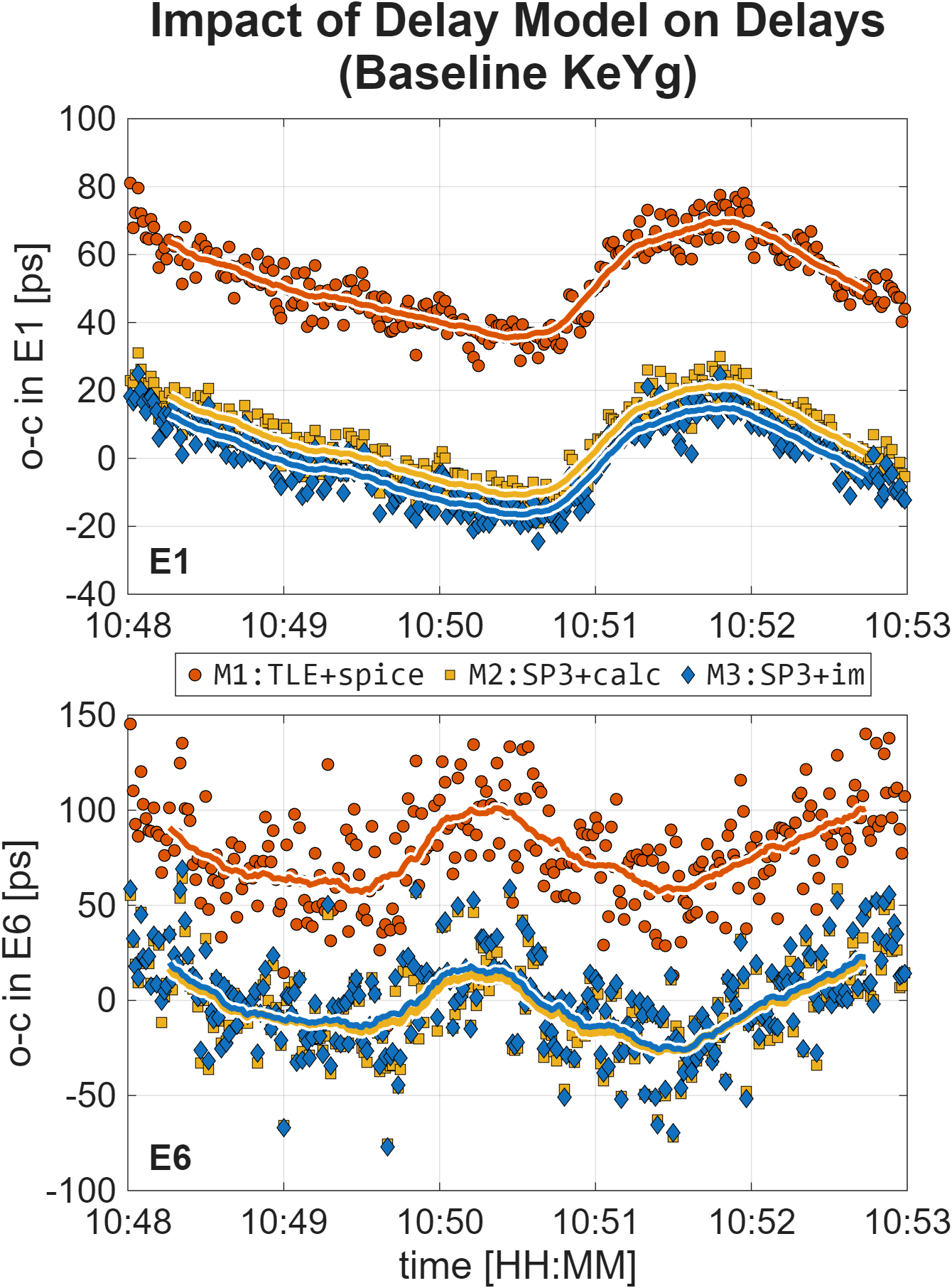}
	\caption{Impact of the delay model on the total delays. Compared are the observed-computed values for a 5-min scan (scan \#4 of experiment A1) when using three different near-field delay model implementations: Duev in \texttt{difxcalc} based on TLEs (\texttt{M1:TLE+spice}), Duev in \texttt{difxcalc} based on SP3 files (\texttt{M2:SP3+calc}) and Klioner in VieVs based on SP3 files (\texttt{M3:SP3+im}). A constant offset and rate were derived from the observed-computed values of the Klioner model and removed from all three solutions. For each solution, a smoothed line using a moving average with a 30-s window is plotted on top of the data.}
	\label{fig:comp_delay_models_1s}
\end{figure}

To investigate the impact of the delay model implementation used in correlation on the delay observables, the 5-min scan (scan \#4 of session A1) was divided into 300 1-s scans. Figure~\ref{fig:comp_delay_models_1s} visualises the observed-minus-computed values of the delays on the Katherine-Yarragadee baseline in the E1 and E6 frequency bands for all three tested near-field delay model implementations. A rate and offset were fit to the delays derived from the \texttt{M3:SP3+im} delay model and removed from all three solutions. A moving average with a 30-s window was generated for each set of residuals.

The data in the figure contains the signal which was also described in Sect.~\ref{sec:04_comp_rec_frg} and shown in Fig.~\ref{fig:moving_average_example}. The signal is apparent in each set of residuals for all three tested delay model implementations in both the E1 and E6 frequency bands. The fact that the course of the signal is similar rules out the delay models as the potential source of the systematic variation in the data.
The differences in the solutions of the individual delay models are mostly characterised by offsets in the residuals. In the E1 frequency band, the average offset between the residuals of the \texttt{M1:TLE+spice} and \texttt{M3:SP3+im} implementations is 53~ps, while it values only about 3~ps between the residuals of \texttt{M2:SP3+calc} and \texttt{M3:SP3+im}.
In the E6 frequency band, the average offsets between the implementations are 79~ps between \texttt{M1:TLE+spice} and \texttt{M3:SP3+im} and -2~ps between the residuals of \texttt{M2:SP3+calc} and \texttt{M3:SP3+im}. Although the implementations provide the identical, frequency-independent a priori delay information for both the E1 and E6 frequency bands, the offsets between the models differ from E1 to E6.
We assume that the reason for the offsets and variation in the offsets for different frequency bands is caused by the degree of agreement between the delay in the data and the model.
The residuals demonstrate that the precision of the total delays is hardly affected by the choice of the a priori delay model to the correlator.

\begin{figure*}
	\centering
	\includegraphics[width=\linewidth]{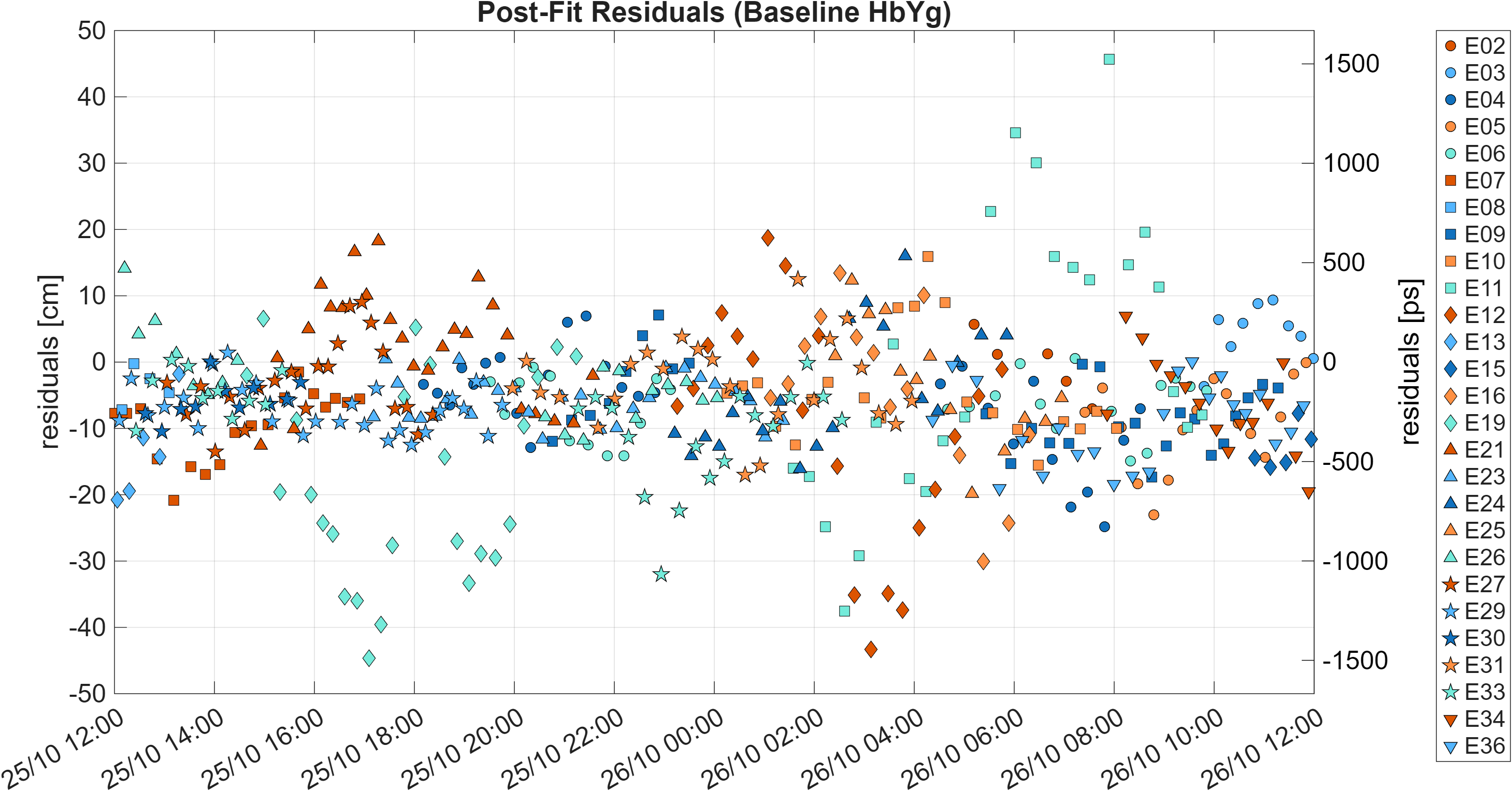}
	\caption{Post-fit residuals for the 8-bit session B2 for the Hobart-Yarragadee baseline. The 24-h session was evaluated in a geodetic analysis using all three baselines from the presented station network with observations in the XX polarisation product. Different markers and colours represent the individual 27~observed Galileo satellites in the session provided as PRN identifiers.}
	\label{fig:postfit_res}
\end{figure*}

As expected, the implementations \texttt{M2:SP3+calc} and \texttt{M3:SP3+im} show closer agreement with each other than with \texttt{M1:TLE+spice}, since both are based on identical orbit information from SP3 files. Furthermore, their smaller residual phase rates underline that SP3 files provide more accurate orbit information than TLEs.
As the \texttt{M3:SP3+im} near-field delay model implementation leads to the highest SNR, a constant amplitude and lowest residual phase delay rate among the three tested approaches, it is considered to be the best performing implementation tested in this study.
Hence, in the previous and following investigations in Sect.~\ref{sec:04_results}, the near-field model implementation \texttt{M3:SP3+im} is utilised for the correlation of the data.

\subsection{Parameter estimation in geodetic analysis}

From the observations of the sessions B1--B4, the coordinates of the VLBI antennas are estimated in a least-squares adjustment using VieVS. 
Due to the geometric configuration of VLBI observations to near-field targets, the normal equations have a full rank, so that datum constraints are not required. Hence, a no-net-translation (NNT) and no-net-rotation (NNR) constraint is not applied.
Besides the station positions, the troposphere is estimated as zenith wet delays every 15~min as piece-wise linear offsets (PWLO) with a relative constraint of 1.5~cm. North-south and east-west gradients are estimated as PWLO every 30~min with relative constraints of 0.05~cm. The clock behaviour is parameterised with a quadratic term and PWLO every 60~min with a relative constraint of 1.3~cm. The positions of the satellites at the time of the scans are fixed to the given a priori orbit information. EOPs are provided as the EOP 20 C04 series\footnote{\url{https://hpiers.obspm.fr/iers/eop/eopc04_20_v2/}} and are also not estimated.

Figure~\ref{fig:postfit_res} shows the typical post-fit residuals for the session B2 on the Hobart-Yarragadee baseline. The residuals for each satellite are not scattered as noise but show strong systematic effects, indicating remaining effects in the data which were not accounted for in the processing or analysis.
Observations to the three satellites E11, E12 and E19 create the largest residuals among the 27~observed Galileo satellites in the session with maximum values of about $\pm$45~cm ($\pm$1500~ps). 
The observations to the remaining satellites have residuals ranging from about -30~cm (-1000~ps) to 20~cm (667~ps).

\begin{table}[b!]
	\centering
	\begin{tabular*}{\columnwidth}{@{\extracolsep{\fill}} cc|cccc}
		\toprule
		& & B1 & B2 & B3 & B4 \\ \midrule
		\multicolumn{1}{c}{\multirow{2}{2.2em}{WRMS}} & [cm] & 20.5 & 12.1 & 12.1 & 11.7 \\
		                                              & [ps] & 683 & 404 & 402 & 390 \\
		\bottomrule
	\end{tabular*}
	\caption{Weighted root mean square (WRMS) values of the post-fit residuals for observing sessions B1--B4. Session B1 was recorded with 2-bit sampling, and fringe fitting was performed over the full 32~MHz channel bandwidth. Sessions B2--B4 were recorded with 8-bit sampling, and fringe fitting was carried out using empirically optimised frequency ranges within channels of 64~MHz bandwidth.}
	\label{tab:wrms_residuals}
\end{table}

The post-fit residuals shown in Fig.~\ref{fig:postfit_res} are representative for the size of residuals for all baselines of the 8-bit observing session B2--B4. In contrast, the residuals for the 2-bit session B1 are almost twice the size. Table~\ref{tab:wrms_residuals} compares the weighted root mean square (WRMS) values of the post-fit residuals for the 24-h observing sessions B1--B4. For the sessions B2--B4, the WRMS values are on the order of 12~cm, the equivalent of about 400~ps. With a WRMS value of 20.5~cm (683~ps), session B1 performs significantly worse. As it is the key feature distinguishing the B1 from the B2--B4 sessions, the smaller sampling bit-depth of 2-bit is the likely cause of the larger post-fit residuals.

\begin{figure}[t!]
	\centering
	\includegraphics[width=\linewidth]{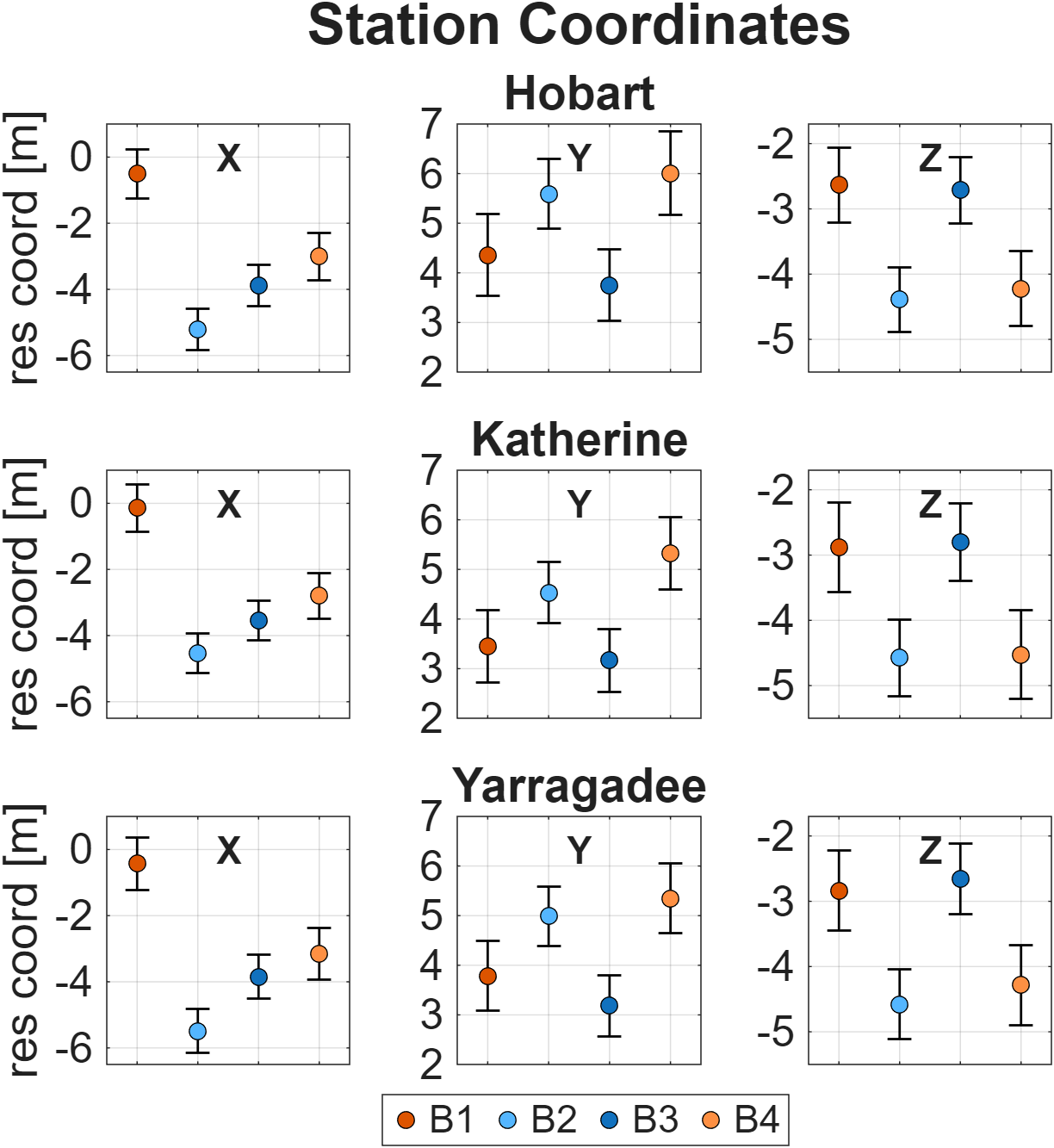}
	\caption{Estimated residual station coordinates for Hobart, Katherine and Yarragadee with respect to the a priori station positions resulting from the geodetic analysis of the four 24-h sessions B1--B4. The error bars represent the associated formal errors.}
	\label{fig:station_positions}
\end{figure}

\begin{figure}[b!]
	\centering
	\includegraphics[width=\linewidth]{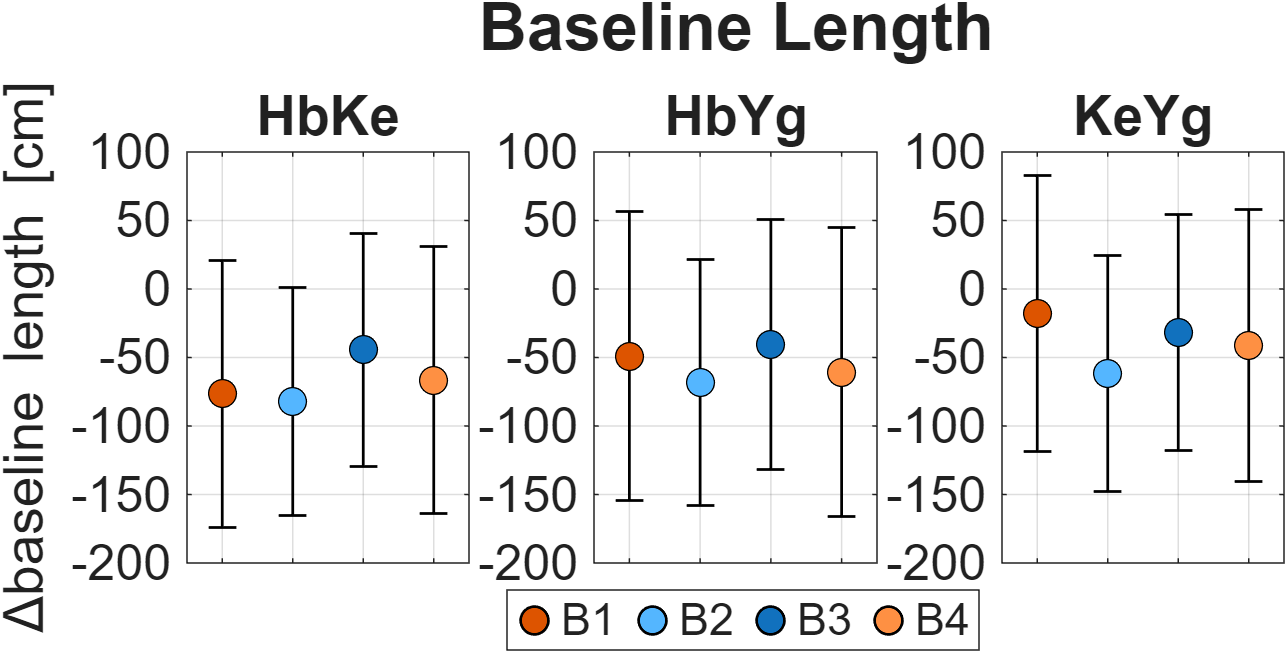}
	\caption{Estimated residual baseline length for Hobart-Katherine, Hobart-Yarragadee and Katherine-Yarragadee with respect to the a priori baseline lengths. The error bars represent the associated formal errors.}
	\label{fig:baseline_length}
\end{figure}

\begin{figure}[t!]
	\centering
	\includegraphics[width=\linewidth]{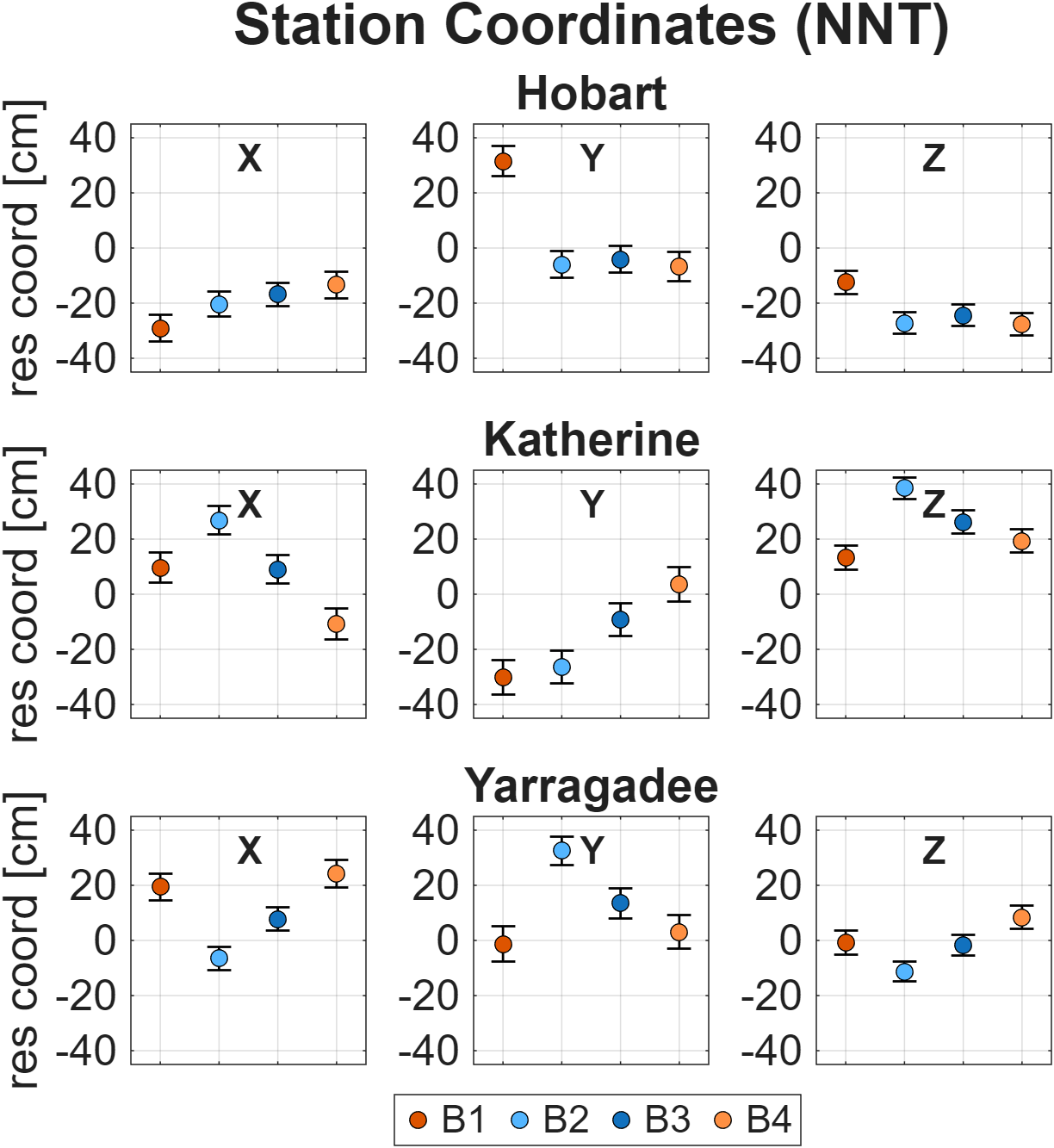}
	\caption{Estimated residual station coordinates for Hobart, Katherine and Yarragadee with respect to the a priori station positions resulting from a datum-restricted geodetic analysis of the four sessions B1--B4, in which an NNT constraint is applied to the estimation of station coordinates. The error bars represent the associated formal errors.}
	\label{fig:station_positions_nnt}
\end{figure}

Figure~\ref{fig:station_positions} shows the estimated residuals to the a priori station coordinates for Hobart, Katherine and Yarragadee. The residuals value up to $\pm$6~metres for individual coordinates. Overall, the coordinates show clear systematic effects. For each observing session, the network experiences significant translations with regard to the a priori position. For example, for session B3, the x-coordinates for all three stations are shifted by about $-$4~m, the y-coordinates by about 3.5~m and the z-coordinates by about $-$3.2~m. As the coordinate residuals are similar in value across all three stations for individual sessions, it can be concluded that these offsets are mainly caused by a translation of the network. To investigate the change in the geometric configuration of the estimated network with respect to the a priori coordinates, the change in the baseline lengths are visualised in Fig.~\ref{fig:baseline_length}. The estimated baselines differ at the sub-metre level. It is evident that, compared to the a priori values, the baselines are estimated methodically shorter. For a specific baseline, the four solutions from sessions B1--B4 vary by about 40~cm. For example, on the Hobart-Katherine baseline, the largest difference is about $-$85~cm for session B2 and the smallest difference is about $-$45~cm.

The estimated station coordinates indicate that the network mainly experiences a translation with respect to the a priori coordinates. To investigate the rotation of the network geometry, an NNT constraint is applied to the estimation of the station coordinates. The resulting residual coordinates are visualised in Fig.~\ref{fig:station_positions_nnt}. They value at maximum up to $\pm$40~cm. As expected, this is a significant decrease compared to the unconstrained solution. The systematic behaviour of the constrained residual coordinates indicate that a significant portion is caused by a remaining rotation of the network relative to the orientation of the a priori station positions.

\section{Discussion}\label{sec:05_dicussion}

In this study, we demonstrate the capability of the AuScope VLBI array to conduct observations of GNSS satellites in L-band at critical scale. This capability is evidenced by an unprecedented comprehensiveness of observations comprising of test experiments as well as four full-scale 24-h sessions targeting Galileo satellites. Given the current lack of practical VLBI observations to satellites within the research community, the presented experiments provide the potential to support the development and optimisation of the VLBI technique for satellite tracking and lay the groundwork for a full implementation into routine operations and procedures.

The results indicate that many refinements are still required in the data processing. The analysis reveals the presence of unmodelled signals in the observations. These signals differ between the E1 and E6 frequency bands, despite both being processed using identical components in the L-band signal chain. In addition, systematic effects are observed when comparing the XX and YY polarisation products. The signals value up to several tens of picoseconds. Whether these effects come from the correlation process, which is not specifically designed to process modulated navigation signals, or are inadvertently introduced elsewhere in the L-band signal chain remains to be investigated.

The derived delay observables exhibit estimated precisions on the order of a few picoseconds in the E1 band and better than 100~picoseconds in the E6 band for 1-s integration times. A comparison between the 2-bit and 8-bit recording modes show that the improved precision achieved with 8-bit sampling justifies the higher recording rates. Depending on the frequency band, baseline and fringe mode, precision improvements of up to a factor of two are observed. These results demonstrate that the L-band signal chain can be used to accurately record the strong, narrow-band navigation signals transmitted by Galileo satellites. It is important to underline, that the presented observations are not in the typical functional range of geodetic VLBI. The navigation signals in L-band are outside the nominal operating range of the receiver and are tremendously stronger than the faint signals from natural geodetic radio sources. The results demonstrate the observations can be performed with reasonably low levels of distortion, compression and saturation effects.

The adopted stepwise tracking strategy proved suitable for tracking Galileo satellites with the presented antenna network. However, the high precision of the E1 delay observables reveals a tracking-related signal in the data. A distinct striped pattern with an amplitude of about $\pm$20~ps is present in the residuals of scans using tracking update intervals of 30~s and 20~s. No comparable pattern is observed for update intervals of 10~s and 5~s. Nevertheless, it cannot be ruled out that the frequent acceleration and deceleration of the antenna steering contributed to an increase in the noise. For future experiments, a smooth and continuous tracking is therefore preferred.

An investigation of the interferometric delay model implementations shows their impact on the resulting delay observables. Significant differences are observed in terms of SNR, residual phase rate and residual delays. While the two SP3-based implementations differ by only a few picoseconds, the TLE-based implementation shows discrepancies on the order of several tens of picoseconds. The residual delays of all three implementations reveal the same underlying signal, indicating that inaccuracies in the interferometric delay models are not the cause of the unmodelled signals in the data. Instead of comparing the resulting delay observables, one could directly compare the absolute values of the input delay models. However, the focus of this work is not to improve the agreement of the delay models but rather to assess currently available methods for deriving the interferometric models to the correlator for near-field sources and evaluate their performance. It is important to point out that the investigation did not compare the underlying theoretical delay models but their implementation.

The post-fit residuals reach values of several decimetres and exhibit systematic satellite-specific behaviour, further confirming the presence of unmodelled signals in the data. One possible source for some of the residuals is the application of phase centre offsets for the Galileo satellites. The utilised SP3 files refer to the satellites' centre of mass rather than their transmit antenna phase centres. The offset would cause a periodic signal in the delays. \citet{McCallum2017jog} report a maximum magnitude of this effect of about 100~ps when observing GNSS satellites. This is significantly smaller than the presented post-fit residuals, which reach values of more than $\pm$1000~ps.

The estimated station coordinate offsets are on the order of several metres. The network primarily experiences a translational offset relative to the a priori coordinates. The baseline lengths are estimated systematically shorter by several decimetres. When restricting the estimated station coordinates by applying an NNT constraint, the station coordinate offsets improve by approximately one order of magnitude, decreasing from several metres to several decimetres.

In addition to the three 12-m antennas of the AuScope VLBI array, the 30-m antenna in Ceduna \citep{McCulloch2005aj} in South Australia participated in the observing session B1. The Ceduna radio telescope is equipped with an astronomical L-band receiver. During session B1, four 10~min scans were conducted to two of the strongest natural geodetic radio sources in L-band visible from the southern hemisphere, 3C273 and 3C279. While no fringes were detected for baselines between the 12-m antennas, weak fringes with SNRs of approximately 10 were detected on baselines to Ceduna. However, fringe detections were not obtained for all polarisation products. Consequently, no manual phase calibration was applied.

Although the metre-level results obtained in this study are approximately three orders of magnitude larger than the millimetre-level accuracy targeted by dedicated missions such as Genesis, this study presents - for the first time - a measured tie between the Galileo (GNSS) and VLBI frames. The empirically determined measurement precision is significantly better than the final frame-tie accuracy, underlining that systematic effects still dominate the solution. The precision of the observables is suitable to investigate these systematic effects in detail. We therefore conclude that these observations provide a valuable test bed for the further development of the technique of VLBI observations of near-field satellites.

\section{Outlook}\label{sec:06_outlook}

For the antennas and observation targets employed in this study, tracking update intervals of 10~s proved to be adequate. However, for the upcoming Genesis mission, substantially shorter tracking update intervals will be required to keep the satellite within the antenna main beam. The stepwise tracking approach used in this work is not suitable for such requirements, as it risks overloading the antenna Field System due to the high rate of positioning updates. While many modern VGOS antennas are capable of continuous satellites tracking, it will be necessary to develop appropriate interfaces for antennas that do not provide this capability, including those used in the present study.

As demonstrated in this work, the VLBI delay depends on a multitude of configuration options across the tracking, observation, correlation and fringe fitting stages. While this study only addressed parts of these topics, further research is required. Some key areas of future research include the coherent combination of polarisation products, phase calibration, ionospheric calibration and determination of multi-band delays.

The presented observing sessions are considered highly experimental. To ensure the success of future co-location satellites missions and to make optimal use of their limited operational lifetime, observations of near-field targets require further methodological and technical development. The developed work flow and observations provide a foundation for improved satellite observations in the future. To assist the research community in these developments, we make the recorded data publicly available.

%
%
%

\backmatter

\bmhead{Acknowledgements}
We would like to thank Dr Warren Hankey for handling the data transfer from the respective VLBI sites to the correlator location at the Mount Pleasant Observatory.

\bmhead{Author Contributions}
DS, LM and JM designed the research. DS planned, coordinated and conducted the observing sessions, processed the data and performed the analysis. JM and DS installed and tested the newly implemented L-band signal chain. DS performed the correlation and fringe fitting of the data, for which JM and TM provided a valuable contribution. DS prepared the dataset for publication and drafted the manuscript. All authors participated in discussions throughout the data collection, processing and analysis process, contributed to the interpretation of the results and reviewed the manuscript.

\bmhead{Funding}
The AuScope VLBI project is managed by the University of Tasmania, contracted through Geoscience Australia. This work was supported by the Australian Research Council (DE180100245).

\bmhead{Data availability}
Due to the large data volumes, long-term archiving of all raw VLBI observation recordings is not feasible. We therefore retain only a selected subset of datasets. The raw VLBI recordings of datasets A1 and B2 are stored locally at the Mount Pleasant Observatory and are available from the corresponding author upon reasonable request. Furthermore, the correlation output of session B2 (SWIN format) as well as the ionosphere-free delay observables of the four 24-h observing sessions in VSO format (with the intention to upgrade to vgosDB format once available) have been deposited in a public repository and are accessible at: \url{https://doi.org/10.60623/odna57ut}.

\section*{Declarations}

\bmhead{Conflict of Interest}
The authors have no conflict of interest to declare.


\bibliography{sn-bibliography}

\end{document}